\documentstyle[12pt]{article}
\begin{document}
\begin{titlepage}
\hskip 12cm \vbox{\hbox{BUDKERINP/98-55} \hbox{July 1998}}
\vskip 0.3cm
\centerline{\bf BFKL NEWS$^{~\ast}$}
\centerline{  V.S. Fadin$^{\dagger}$}
\centerline{\sl Budker Institute for Nuclear Physics,}
\centerline{\sl Novosibirsk State University, 630090
Novosibirsk, Russia}
\vskip 2cm
\begin{abstract}
I discuss  radiative corrections to the  BFKL equation for high
energy cross sections in  perturbative QCD. Due to the gluon
Reggeization  in the
next-to-leading $\ln s$  approximation, the form of the BFKL
equation
remains unchanged and the corrections to the BFKL kernel are expressed
in terms of the
two-loop contribution to the gluon Regge trajectory, the one-loop correction
to the
Reggeon-Reggeon-gluon vertex and the contributions from two-gluon and
quark-antiquark production in Reggeon-Reggeon collisions.  I present the
results
of the calculation of the BFKL kernel
in the next-to leading logarithmic approximation, the estimate of the
Pomeron shift and   the
next-to-leading contribution to the anomalous dimensions of twist-2
operators
near $j=1$.
\end{abstract}

\vskip .5cm
\hrule
\vskip.3cm
\noindent
$^{\ast}${Talk given at "LISHEP98", LAFEX school on high energy physics,
February 14-21, Rio de Janeiro, Brazil, 1998}
\vfill
$ \begin{array}{ll}
^{\dagger}\mbox{{\it email address:}} &
 \mbox{FADIN~@INP.NSK.SU}\\
\end{array}
$
\end{titlepage}

\vskip 0.5cm


\section
{Introduction}

The BFKL equation \cite{FKL} is very popular now, mainly due to recent
experimental
results on deep inelastic scattering of electrons on protons obtained
at
HERA \cite{H1}, which show growth of the gluon density in the proton
with
decreasing of the fraction  of the proton momentum carried by  gluon.
It can be used together with the DGLAP evolution equation \cite{GL} for
the description of structure functions for the deep inelastic $ep$
scattering at small values of the Bjorken variable $x$
(see, for instance, \cite{AKMS} and references therein).
The equation was derived for
scattering amplitudes in QCD at high energies $\sqrt{s}$ and fixed
momentum
transfer $\sqrt{-t}$ in the leading logarithmic
approximation (LLA) which means collection of
all terms of the type $[\alpha_s\ln s]^n$. This approximation leads to
a sharp increase of cross sections with c.m.s. energy $%
\sqrt{s}$. In fact, calculated in LLA,  the total cross section
${\sigma}^{LLA}_{tot}$ grows at large c.m.s.
energies as a power of $s$:
\begin{equation}
{\sigma}^{LLA}_{tot}\sim {s^{\omega^B_{P}}\over {\sqrt{\ln{s}}}}~,
\label{a1}
\end{equation}
where $\omega^B_{P}$ is the LLA position of the most right
singularity in the complex momentum plane  of the  $t$-channel partial
wave with vacuum quantum numbers  (Pomeron singularity),
given by
\begin{equation}
{\omega^B_{P}} = {g^2\over \pi^2}N{\ln{2}}
\label{a2}
\end{equation}
for the gauge group SU({N}) ($N=3$ for QCD) with gauge coupling
constant $g$
($\alpha_s={g^2\over 4\pi}$).
Therefore, the Froissart bound $\sigma_{tot} < const(\ln s)^2$ is
violated in LLA. The reason of the violation is that the
$s$-channel unitarity constraints for scattering amplitudes are not
fulfilled in this approximation. The problem of
unitarization  of LLA results is extremely important from the
theoretical point of view. It is concerned in a lot of papers (see,
for
example, \cite{MLBCN} and references therein).

 The violation of the Froissart bound means that
 LLA can not be applied at asymptotically large energies. But in the
region of energies
accessible for modern experiments it seems that
the most important disadvantage of LLA is that neither the scale of $s$
nor the argument of the running coupling constant $\alpha_s$ are
fixed in this approximation. These uncertainties diminish the
predictive power of LLA, permitting to change
strongly numerical results by changing the scales. From the  practical
point
of view, since the results of LLA are applied to the small $x$
phenomenology, it is extremely
important to remove these uncertainties. Another important problem is
the determination of
the region of energies and
momentum transfers where  LLA could be applicable. To solve these
problems we have to know
radiative corrections to  LLA.

 Therefore, the radiative corrections to the BFKL equation
are very important, as they give the possibility to fix the argument
of the running
coupling, to define the scale of energy and to determine
the region of   applicability of the results obtained.
 My talk is devoted to these radiative corrections.

 The outline of the talk is the following. In Section 2, I remind the
derivation of the BFKL equation in LLA and the solution of this equation.
In Section 3, I discuss the general form of corrections
in the next-to-leading logarithmic approximation (NLLA) and present the
two-loop correction to the gluon Regge trajectory and the contributions
to the NLLA  kernel from the  one-loop correction
to the one-gluon production and from
the two-gluon and quark-antiquark pair production in the
Reggeon-Reggeon collisions.
In Section 4,  all the contributions are collected together,
the cancellation of infrared singularities is performed,  the estimate of
the shift of the Pomeron intercept  and
the next-to-leading contributions to anomalous dimensions of twist-2
operators near $j=1$ are presented.

\section{The BFKL equation in LLA}
Despite the fact that the BFKL equation was obtained  more
than 20 years ago, till now a simple derivation of this equation does
not
exist, though attempts to do it continue (see, for examlpe,
\cite{JKL}).
In the original derivation
\cite{FKL}
the key role was played by the gluon
Reggeization in QCD \cite{GST,L0}. In fact, the derivation can be
performed without
large
difficulties if we adopt the Reggeization. It is worthwhile to
stress here, that the gluon Reggeization in QCD means something more than
usually assumed. Namely, it means not merely that
there is the Reggeon with gluon quantum numbers, negative signature
and trajectory
\begin{equation}
\label{a3}
j(t) = 1+\omega(t)~
\end{equation}
passing through 1 at $t=0$, but also that only this Reggeon gives the
leading contribution in
each order of the perturbation theory  to
amplitudes with the gluon quantum numbers in channels with fixed momentum
transfers.  Due to this property it is not
difficult to
calculate the leading contributions to the imaginary parts of  elastic
scattering amplitudes with arbitrary quantum numbers in  $t$-channel
at large
$s$ and fixed
$t$ using the unitarity
condition.  Full amplitudes are  easily restored
through their imaginary
parts. The BFKL equation emerges from the representation of the
amplitudes in the particular case of the forward scattering with
vacuum quantum numbers in
$t$ - channel.

For the elastic scattering process $A+B \rightarrow A'+B'$ the gluon
Reggeization  means that at large $s$ and fixed $t$, with
\begin{equation}
s=(p_A+p_B)^2,\,\,\, t=q^2,\,\,\, q=p_A-p_{A'}\, ,
\label{a4}
\end{equation}
the amplitudes with gluon quantum numbers in $t$-channel have the
factorized form
\begin{equation}
\left({{\cal A}_8}\right)^{A'B'}_{AB} = \Gamma^i_{A'A}{s\over
t}
\left[\left({s\over -t}\right)^{\omega(t)}+\left({-s\over
-t}\right)^
{\omega(t)}\right]\Gamma^i_{B'B}~,
\label{a5}
\end{equation}
where $\Gamma^i_{A'A}$ are the particle-particle-Reggeon (PPR)
vertices.
In LLA for the deviation of the gluon trajectory from 1 we have \cite{L0}
\begin{equation}
\omega(t) = \omega^{(1)}(t) = {g^2t\over
{(2{\pi})}^{(D-1)}}\frac{N}{2}
\int{d^{D-2}k_\perp\over k_\perp^2{(q-k)}_{\perp}^2}~,
\label{a6}
\end{equation}
where $t=q^2\approx q_\perp^2$ and
$D=4+2\epsilon$ is the space-time dimension. A non-zero
$\epsilon$ is
introduced to regularize Feynman integrals. The integration in
Eq.(\ref{a6})
is performed over  ($D-2$)-dimensional momenta orthogonal to the
initial
particle momentum plane. The PPR vertices can be presented as
\begin{equation}
\Gamma^{i}_{A'A} = g\langle A'| T^i| A \rangle
\Gamma_{A'A}~,
\label{a7}
\end{equation}
where $\langle A'| T^i| A \rangle$ stands for a matrix element of the
colour group generator in the corresponding representation (i.e.
fundamental
for quarks and adjoint for gluons). In LLA  the helicities $\lambda_{P}$ of
each of the scattered particles P are conserved, so
that in the helicity basis we have
\begin{equation}
\Gamma_{A'A} =\Gamma^{(0)}_{A'A}= \delta_{{\lambda_{A'}}{\lambda_A}}~.
\label{a8}
\end{equation}

The $s$ - channel unitarity relation for the  imaginary part of the  elastic
scattering amplitude ${\cal A}^{A'B'}_{AB}$ can be presented as
\begin{equation}
Im_s {\cal A}^{A'B'}_{AB}=\frac 1{2}\sum_{n=0}^\infty \sum_{\{f\}}\int
{\cal A}^{\tilde A  \tilde B +n}_{AB}
\left({\cal A}^{\tilde A  \tilde B +n}_{A'B'}\right)^*
d\Phi _{\tilde A \tilde B +n},
\label{a9}
\end{equation}
where $Im_s$ stands for the $s$ - channel imaginary part,
 ${\cal A}^{\tilde A \tilde B+n}_{AB}$  is the
amplitude of production  of $n$ particles with momenta $k_i,$   $i=1,....n$
in the process $A+B \rightarrow \tilde A + \tilde B +n,\,\,\,$
  $\sum_{\{f\}}$
means sum over the discrete
quantum numbers
of the final particles in this process, $d\Phi_{\tilde A, \tilde B +n}$ is
the element of the final particle phase space.
We admit all particles to have non zero masses (reserving the possibility
to consider each of them as a group of particles) and use light-cone vectors
$p_1$ and $p_2$ such that momenta of the initial
particles $A$ and
$B$ are equal  $p_A=p_1+\left( m_A^2/s\right) p_2$ and
$p_B=p_2+\left(m_B^2/s\right)p_1$ respectively  and $s=2(p_1p_2).$  Using the Sudakov
decomposition

\begin{equation}
k_i=\beta _i p_1+\alpha _i p_2 +k_{i\perp }\, , \,\,\,s\alpha _i\beta
_i=k_i^2 -k_{i\perp }^2=k_i^2 +{\vec k}_i^2,
\label{a10}
\end{equation}
where  the vector sign
denotes (here and below) transverse momenta,
we obtain
\[
d\Phi _{\tilde A, \tilde B +n}
=\frac{2}{s}(2\pi )^D\delta (1+\frac{m_A^2}%
s-\sum_{i=0}^{n+1}\alpha _i)\delta (1+\frac{m_B^2}s-\sum_{i=0}^{n+1}\beta
_i)
\]
\begin{equation}
\times\delta^{D-2} (\sum_{i=0}^{n+1}k_{i\perp })\frac{d\beta _{n+1}}{2\beta _{n+1}}\frac{d\alpha _0}{2\alpha _0}\prod_{i=1}^n\frac{d\beta _i}
{2\beta _i}\prod_{i=0}^{n+1}\frac{d^{D-2}k_{i\perp }}{(2\pi )^{D-1}},
\label{a11}
\end{equation}
with the denotations
\begin{equation}
p_{\tilde A} = k_0, \,\,\, p_{\tilde B} = k_{n+1}.
\label{a12}
\end{equation}

In the unitarity condition (\ref{a9}) the contribution of order $s$,
 which we are
interested in,  is given by the region of limited (not growing with $s$)
transverse momenta of
produced particles. Only this region is considered in the following.
Large logarithms  come from integration over  longitudinal momenta of the
produced particles.
Therefore in LLA, where production of each additional particle must  give
the large
logarithm ($\ln s$),  they are produced in the multi-Regge kinematics (MRK).
By definition, in this kinematics their transverse
momenta are limited and their Sudakov variables
$\alpha _i$ and $\beta _i$ , $i=0\div n+1,$ are strongly
ordered (in another words, the produced particles are strongly ordered in
the rapidity space).
Let us take, for definiteness, that
\begin{equation}
\alpha _{n+1} \gg \alpha _n\gg \alpha _{n-1}....\gg \alpha _0,
\,\,\,\,
\beta _0 \gg \beta _1\gg \beta _2....\gg \beta _{n+1}.
\label{a13}
\end{equation}
In this case the $\delta -$ functions in Eq. (\ref{a11}) give us
\begin{equation}
\alpha _{n+1}\simeq 1,\,\,\,\beta _0\simeq 1
\label{a14}
\end{equation}
and therefore
\begin{equation}
\alpha _0\simeq \frac{{{\vec p}_{\tilde A}}^{~2}+m_{\tilde A}^2}s, \,\,\,\,\,
\beta _{n+1}\simeq
\frac{{{\vec p}
_{\tilde B}}^{~2}+m_{\tilde B}^2}s.
\label{a15}
\end{equation}
In  MRK the squared  invariant masses $s_{ij} =(k_i+k_j)^2$ of any pair of
 produced
particles $i$ and $j$ are large. In order to obtain the large logarithm from
the integration
over $\beta_i$ for each produced particle in the phase space (\ref{a11}),
 the amplitudes in the
r.h.s. of the unitarity relation (\ref{a9}) must not decrease with the
growth of the invariant
masses. It is possible only in the case where there are exchanges of
vector particles
(i.e. gluons) in all channels with momentum transfers $q_i$, \,\,\,
 $i=1\div n+1$, whith
\begin{equation}
q_i = p_A -\sum_{j=0}^{i-1}k_j =-(p_B-\sum_{l=i}^{n+1}k_l)
\simeq \beta _i p_1-\alpha _{i-1} p_2 -\sum_{j=0}^{i-1}k_{j\perp };\,\,\,\,\,
q_i^2 \simeq q_{i\perp}^2 = -{\vec q}_{i}^{~2}.
\label{a16}
\end{equation}
Due to the gluon Reggeization the amplitudes of such processes in LLA have simple
multi-Regge form:
\begin{equation}
{\cal A}^{\tilde A  \tilde B +n}_{AB}=2s\Gamma_{\tilde A A}^{c_1}
\left( \prod_{i=1}^n \gamma _{c_ic_{i+1}}^{P_i}(q_i,q_{i+1})
\left(\frac{s_i}{s_R}\right) ^{\omega(t_i)}\frac 1{t_i}\right)
\frac 1{t_{n+1}}
\left( \frac{s_{n+1}}{s_R}\right) ^{\omega(t_{n+1})}
\Gamma _{\tilde B B}^{c_{n+1}} ,
\label{a17}
\end{equation}
where $s_R$ is some energy scale, which is irrelevant in LLA,
\begin{equation}B
s_i=(k_{i-1}+k_i)^2\simeq s\beta _{i-1}\alpha _i=\frac{\beta _{i-1}}{\beta _i}
(\overrightarrow{k_i}^2 +k_i^2), \,\,\,
t_i=q_i^2\simeq -\overrightarrow{q_i}^2,
\label{a18}
\end{equation}
$\omega(t)$  and $\Gamma _{P^{^{\prime }}P}^{a}$ are the gluon
Regge trajectory and the PPR
 vertices given by
Eqs.(\ref{a6}) and (\ref{a7}), (\ref{a8}) respectively;
$\gamma _{c_ic_{i+1}}^{P_i}(q_i,q_{i+1}) $ are the effective
vertices of production of particles $P_i$ with momenta $q_i-q_{i+1}$ in
collisions of Reggeons
(i.e. Reggeized gluons) with momenta $q_i$ and $-q_{i+1}$ and colour indices
$c_i$ and $c_{i+1}$
correspondingly. In LLA all produced particles $P_i$ must be gluons;
therefore, the masses
of produced particles are equal to zero. The Reggeon-
Reggeon-gluon  (RRG) vertex  has the form \cite{FKL}

\begin{equation}
\gamma_{c_ic{i+1}}^{G_i}(q_i,q_{i+1})
=gT_{c_ic_{i+1}}^{d_i}e^*_{\mu}(k_i)C^\mu (q_{i+1},q_i),
\label{a19}
\end{equation}
where $T_{c_ic_{i+1}}^{d_i}$ are the  matrix elements of  the
$SU(N)$ group generators
in the
adjoint representation, $d_i$ is the colour index of the produced gluon,
$e(k_i)$ its
polarization vector and $k_i=q_i-q_{i+1}$  its momentum;
\begin{equation}
C^{\mu }(q_{i+1},q_i) = -q_i-q_{i+1} +
p_1(\frac{q_i^2}{k_ip_1}+2\frac{k_ip_2}{p_1p_2}) - p_2(\frac{q_{i+1}^2}{k_ip_2}+
2\frac{k_ip_1}{p_1p_2}).
\label{a20}
\end{equation}
The amplitude ${\cal A}^{\tilde A  \tilde B +n}_{A'B'}$ entering  the unitarity relation
(\ref{a9}) can be obtained from Eq. (\ref{a17}) by the substitutions $A\rightarrow A'\, ,
B \rightarrow B'\, , q_i \rightarrow q'_i \equiv q_i-q, \,$ where $q =p_A-p_{A'}
\simeq q_{\perp}$.

Let us introduce the operators $\hat{\cal P}_R$ for projection of
two-gluon colour states in  $t-$ channel in the unitarity
condition (\ref{a9}) on the
irreducible representations $R$ of the colour group and use the decomposition
\begin{equation}
T_{c_ic_{i+1}}^{d_i}(T_{c'_ic'_{i+1}}^{d_i})^* = \sum_R c_R\langle
c_ic'_i|\hat{\cal P}_R |c_{i+1}c'_{i+1} \rangle.
\label{a21}
\end{equation}
We'll be interested in the singlet (vacuum) and antisymmetrical octet (gluon) representations.
For the first of them
\begin{equation}
 \langle c_ic'_i|\hat{\cal P}_0 |c_{i+1}c'_{i+1} \rangle =
 \frac{\delta_{c_ic'_i}\delta_{c_{i+1} c'_{i+1}}} {N^2-1}
\label{a22}
\end{equation}
and for the second
\begin{equation}
 \langle c_ic'_i|\hat{\cal P}_8 |c_{i+1}c'_{i+1} \rangle =
 \frac{f_{ac_ic'_i}f_{ac_{i+1}c'_{i+1}}}{N},
 \label{a23}
\end{equation}
so that one can easily find
\begin{equation}
c_o=N, \,\,\, c_8= \frac{N}{2}.
\label{a24}
\end{equation}
Using the decomposition (\ref{a21}) we obtain from (\ref{a19})
\begin{equation}
\sum_{G_i} \gamma_{c_ic_{i+1}}^{G_i}(q_i,q_{i+1})
(\gamma_{c'_ic'_{i+1}}^{G_i}(q_i,q_{i+1} ))^*
=
\sum_R \langle
c_ic'_i|\hat{\cal P}_R |c_{i+1}c'_{i+1} \rangle 2(2\pi )^{D-1}
{\cal K}_r^{(R)}(\vec q_i,\vec q_{i+1};\vec q ),
\label{a25}
\end{equation}
where the sum is taken over colour and polarization states of the produced
gluon and
\[
{\cal K}_r^{(R)}(\vec q_i,\vec q_{i+1};\vec q )=
-\frac{g^2c_R}{2(2\pi )^{D-1}}
C^{\mu }(q_{i+1},q_i) C_{\mu }(q_{i+1}-q,q_i-q)
\]
\begin{equation}
=
\frac{g^2c_R}{(2\pi )^{D-1}}
\left(\frac{{\vec q}_i^2{(\vec q_{i+1} -\vec q)}^2
+{\vec q}_{i+1}^2{(\vec q_{i} -\vec q)}^2}{({\vec q}_i-{\vec q}_{i+1})^2}
-\vec q^{~2} \right).
\label{a26}
\end{equation}

The decomposition (\ref{a21}) corresponds to the decomposition of
 the elastic scattering  amplitude ${{\cal A}}^{A'B'}_{AB}$ in  (\ref{a9}):
\begin{equation}
{{\cal A}}^{A'B'}_{AB} =\sum_R \left({{\cal A}_R}\right)^{A'B'}_{AB},
\label{a27}
\end{equation}
where ${\cal A}_R$ is the part of the scattering amplitude corresponding to
the definite irreducible representation
$R$ of the colour group in  $t$ -channel. It is convenient to consider
its partial
wave $f_R(\omega, \vec q)^{A'B'}_{AB}$
defined by
\begin{equation}
f_R(\omega, \vec q)^{A'B'}_{AB} = \int_{s_0}^\infty \frac{ds}{s^2}
\left(\frac{s}{s_0}\right)^{-\omega}Im_s\left({{\cal A}_R}\right)^{A'B'}_{AB}.
\label{a28}
\end{equation}
The amplitude itself is expressed trough the partial wave as
\begin{equation}
\left({{\cal A}_R}\right)^{A'B'}_{AB}=
\frac {s}{2\pi }
\int_{\delta -i\infty }^{\delta +i\infty }\frac{d\omega}{\sin(\pi \omega)}
\left(\left(\frac{-s}{s_0}\right)^\omega -\tau
\left(\frac{s}{s_0}\right)^\omega \right)
f_R(\omega, \vec q)^{A'B'}_{AB},
\label{a29}
\end{equation}
where $\tau$ is the signature and coincides with the symmetry of
the representation $R$. For the gluon representation the Born contribution must be
added into the r.h.s. of (29).
The term with $\tau$ takes into account the contribution to the amplitude from  the
$u$ -channel imaginary part.
Pay attention that the only antisymmetrical representation contributing to the decomposition
(\ref{a21}) is the representation with the gluon quantum numbers. Therefore, only for
this representation the contributions of $s$- and
$u$- channel imaginary parts do not cancel each other. It means that in
each order of  perturbation theory the amplitudes with the gluon quantum numbers in
$t_i$ -channels are dominant.

Let us calculate the contribution $f_R^{(n)}(\omega, \vec q)^{A'B'}_{AB}$
into the partial wave coming from production of $n$ gluons.   Using
Eqs.(\ref{a11}), (\ref{a13})-(\ref{a15}), (\ref{a18}), we have
\[
s=\left( \prod_{i=1}^{n+1}s_i\right) \left( \prod_{i=1}^n{\vec k}_i^2\right)
^{-1}= \left( \prod_{i=1}^{n+1}\frac{s_i}{\sqrt{{\vec k}_{i-1}^{~2}
{\vec k}_i^{~2}}}\right) \sqrt{{\vec q}_1^{~2}{\vec q}_{n+1}^{~2}},\,\,\,
\]
\begin{equation}
\frac{ds}sd\Phi _{\tilde A \tilde B +n}=\frac{2\pi }s
\prod_{i=1}^{n+1}\frac{ds_i}{2s_i}\frac{d^{D-2}q_{i\perp }}{(2\pi )^{D-1}}
\label{a30}
\end{equation}
and from Eqs. (\ref{a28}), (\ref{a9}), (\ref{a17}) and (\ref{a25}) we obtain
\[
f_R^{(n)}(\omega, \vec q)^{A'B'}_{AB}=\frac{1}{(2\pi )^{D-2}}
\int\left(\prod_{i=1}^{n+1}\frac{d^{D-2}q_{i\perp }}{\vec q^{~2}_{i}
({\vec q}_{i}-{\vec q})^2}\frac{ds_i}{s_i}%
^{}\left( \frac{s_i}{s_R}\right) ^{\omega(t_i)+\omega(t'_i)}
\left(\frac{s_i}{\sqrt{{\vec k}_{i-1}^{~2}
{\vec k}_i^{~2}}}\right)^{-\omega} \right)
\]
\begin{equation}
\times \left(\frac{s_0}{\sqrt{{\vec q}_1^{~2}{\vec q}_{n+1}^{~2}}}\right)
^{\omega}\sum_{\nu}I_{A'A}^{(R,\nu)}
\left(\prod_{i=1}^n{\cal K}^{(R)}_r(\vec q_i,
\vec q_{i+1};\vec q)\right) I_{B'B}^{(R,\nu)}.
\label{a31}
\end{equation}
The index $\nu$ here enumerates the states in the irreducible
representation $R$, so that
\begin{equation}
\langle c_ic'_i|\hat{\cal P}_R |c_{i+1}c'_{i+1} \rangle =
\sum_{\nu} \langle c_ic'_i|\hat{\cal P}_R |\nu \rangle
\langle \nu|\hat{\cal P}_R |c_{i+1}c'_{i+1} \rangle
\label{a32}
\end{equation}
and
\[
I_{A'A}^{(R,\nu)} = \sum_{{\tilde A}} \Gamma_{\tilde A A}^{c_1}
\left(\Gamma_{\tilde A A'}^{c'_1}\right)^*
\langle c_1c'_1|\hat{\cal P}_R |\nu \rangle
\]
\begin{equation}
I_{B'B}^{(R,\nu)} = \sum_{{\tilde B}}\Gamma_{\tilde B B}^{c_{n+1}}
\left(\Gamma_{\tilde B B'}^{c'_{n+1}}\right)^*
\langle \nu|\hat{\cal P}_R |c_{n+1}c'_{n+1} \rangle.
\label{a33}
\end{equation}
The sum here is taken over the discreet quantum numbers of the states $\tilde A$,
$\tilde B$.

For the singlet representation the index $\nu$ takes only one value, so
that we can omit it and  have
\begin{equation}
\langle c c'|\hat{\cal P}_0|0 \rangle =\frac{\delta_{cc'}}{\sqrt {N^2-1}},
\label{a34}
\end{equation}
whereas  for the antisymmetrical octet (gluon) representation the index
$\nu$ coincide with gluon colour index and
\begin{equation}
\langle c c'|\hat{\cal P}_8|a \rangle =\frac{f_{acc'}}{\sqrt N},
\label{a35}
\end{equation}
where $f_{abc}$ are the structure constants of the colour group.
Therefore, we obtain
\begin{equation}
I_{P'P}^{(0)} =  \frac {g^2 C_P}{\sqrt{N^2-1}} \langle P'| P \rangle
\delta_{{\lambda_{P'}}{\lambda_P}},
\label{a36}
\end{equation}
where $C_P$ is the value of the Casimir operator in corresponding representation,
i.e. $C_P=\frac {N^2-1}{2N}$ for quarks and  $C_P=N$ for gluons;
\begin{equation}
I_{P'P}^{(8, a)} = -ig^2\frac{\sqrt N}{2} \langle P'|T^a| P \rangle
\delta_{{\lambda_{P'}}{\lambda_P}} = -ig\frac{\sqrt N}{2} \Gamma^a_{P'P}\, .
\label{a37}
\end{equation}
The integration over $s_i$ in Eq.(\ref{a31}) is performed from some
fixed (independent from $s$) value to infinity. Note, that  the essential
integration region
in Eq. (\ref{a29}) is $\omega \sim (\ln s)^{-1}$, so that in LLA
we can omit terms of the type $\omega \ln {s_R}, \,\,
\omega \ln {\vec k_i^{~2}}$, as well as $\omega(t_i)\ln {s_R}, \,\,
\omega(t'_i)\ln {s_R}$. It corresponds to the statement that in LLA the
scale of energy is not fixed. Therefore, independently from the lower limit
of the integrations over $s_i$  we obtain
\[
f_R^{(n)}(\omega, \vec q)^{A'B'}_{AB}=\frac{1}{(2\pi )^{D-2}}
\int\prod_{i=1}^{n+1}\frac{d^{D-2}q_{i\perp}}{(\omega -\omega(t_i)
-\omega(t'_i))\vec q_i^{~2}({\vec q}_{i}-{\vec q})^2 }
\]
\begin{equation}
\times \sum_{\nu}I_{A'A}^{(R,\nu)}\left(\prod_{i=1}^n{\cal K}^{(R)}_r(\vec q_i,
\vec q_{i+1};\vec q)\right) I_{B'B}^{(R,\nu)}.
\label{a38}
\end{equation}
Let us present the partial wave in the form
\begin{equation}
f_R(\omega, \vec q)^{A'B'}_{AB}= \frac{1}{(2\pi )^{D-2}}
\int \frac{d^{D-2}q _{A\perp}}{\vec q_A^{~2}({\vec q}_{A}-{\vec q})^2 }
\frac{d^{D-2}q _{B\perp }}{\vec q_B^{~2}({\vec q}_{B}-{\vec q})^2 }
\sum_{\nu}I_{A'A}^{(R,\nu)}
G^{(R)}_{\omega}({\vec q}_{A}, {\vec q}_{B}; {\vec q})
I_{B'B}^{(R,\nu)} \, .
\label{a39}
\end{equation}
Then for the function $G^{(R)}_{\omega}$ (which can be called Green function
for scattering of two Reggeized gluons) we obtain
\[
G^{(R)}_{\omega}({\vec q}_{A}, {\vec q}_{B}; {\vec q})=
\sum_{n=0}^{\infty}\int\left(\prod_{i=1}^{n+1}\frac{d^{D-2}q_{i\perp }}
{ \vec q_i^{~2}({\vec q}_{i}-{\vec q})^2
(\omega -\omega(t_i)-\omega(t'_i)) }\right)
\]
\begin{equation}
\left(\prod_{i=1}^n{\cal K}^{(R)}_r(\vec q_i,
\vec q_{i+1};\vec q)\right){\vec q}_A^{~2}({\vec q}_A-{\vec q})^2
{\vec q}_B^{~2}({\vec q}_B-{\vec q})^2
\delta^{D-2}(q_{1\perp }-q_{A\perp })
\delta^{D-2}(q_{n+1\perp }-q_{B\perp }).
\label{a40}
\end{equation}
It is easy to see, that the sum in the r.h.s. is the perturbative
solution of the equation
\[
\omega G^{(R)}_{\omega}({\vec q}_1, {\vec q}_2; {\vec q}) =
\vec q_1^{~2}({\vec q}_{1}-{\vec q})^2\delta^{(D-2)}
({\vec q}_1- {\vec q}_2)
\]
\begin{equation}
 +\int \frac{d^{D-2}q_{1\perp}^{~\prime}}
{\vec q_1^{~\prime 2}({\vec q}_{1}^{~\prime}-{\vec q})^2 }
{\cal K}^{(R)}({\vec q}_1, {\vec q}_1^{~\prime}; {\vec q})
G^{(R)}_{\omega}({\vec q}_1^{~\prime}, {\vec q}_2; {\vec q}),
\label{a41}
\end{equation}
where the kernel
\begin{equation}
{\cal K}^{(R)}({\vec q}_1, {\vec q}_2; {\vec q}) =(\omega(q_{1\perp}^2)+
\omega((q_1-q)_{\perp}^2))\vec q_1^{~2}({\vec q}_{1}-{\vec q})^2
\delta^{(D-2)}({\vec q}_1- {\vec q}_2)
+{\cal K}^{(R)}_r({\vec q}_1, {\vec q}_2; {\vec q})
\label{a42}
\end{equation}
consists of two parts; the first of them, so called virtual part,
is expressed in terms of the gluon Regge trajectory and the second, related
with the real particle production, is given by Eq.(\ref{a26}).

  Let us consider the partial wave $f_R^{(\nu)}(\omega;
{\vec q}_1,{\vec q})_B^{B'}$ for  the Reggeized gluon scattering off the
 particle $B$ defined as
\begin{equation}
f_R^{(\nu)}(\omega; {\vec q}_1,{\vec q})_B^{B'} =\int d^{D-2}q _{B
\perp }
G^{(R)}_{\omega}({\vec q}_{1}, {\vec q}_{B}; {\vec q})
\frac{I_{B'B}^{(R,\nu)}}{\vec q_B^{~2}({\vec q}_{B}-{\vec q})^2},
\label{a43}
\end{equation}
where the impact factor $I_{B'B}^{(R,\nu)}$ is given by Eq.(\ref{a33}).
This partial wave satisfies the equation
\begin{equation}
\omega f_R^{(\nu)}(\omega; {\vec q}_1,{\vec q})_B^{B'}
 = I_{B'B}^{(R,\nu)}
+\int \frac{d^{D-2}q_{1\perp}^{~\prime}}
{{\vec q_1^{~\prime 2}({\vec q}_{1}^{~\prime}-{\vec q})^2 } }
{\cal K}^{(R)}({\vec q}_1, {\vec q'}_1; {\vec q})
f_R^{(\nu)}(\omega; {\vec q'}_1,{\vec q})_B^{B'}.
\label{a44}
\end{equation}
It is easy to see that for the case of the gluon representation the
solution of this equation is
\begin{equation}
f_8^{a}\left(\omega; {\vec q}_1,{\vec q} )\right)_B ^{B'}=
\frac{I_{B'B}^{(8, a)}}{\omega - \omega(q_{\perp}^2)}.
\label{a45}
\end{equation}
Therefore Eqs. (\ref{a39}), (\ref{a43})-(\ref{a45}) give us
\[
f_8(\omega, {\vec q})_{AB}^{A'B'}=I_{A'A}^{(8, a)}\frac{1}{(2\pi )^{D-2}}
\int \frac{d^{D-2}q _{1\perp}}{\vec q^{~2}_{1}({\vec q}_{1}-{\vec q})^2}
I_{B'B}^{(8,a)}\frac{1}{\omega - \omega(q_{\perp}^2)}
\]
\begin{equation}
=\Gamma_{A'A}^{a}\frac{-\pi\omega(t)}{t(\omega - \omega(t))}\Gamma_{B'B}^{a}.
\label{a46}
\end{equation}
Exactly the same result one gets substituting Eq.(\ref{a5}) in Eq.(\ref{a28}).
So, we have a kind of "bootstrap": we input the existence of the
Reggeized gluon and obtained it as the solution of the equation derived
from the unitarity.

Strictly speaking, this "bootstrap", although being very impressive and
meaningful,  can not be considered as a rigorous proof of the gluon
Reggeization. For such proof we have to reproduce not only the form
(\ref{a5}) of the elastic amplitudes, but also the Reggeized form
(\ref{a17}) for all inelastic amplitudes. It was actually done
 in LLA  in Ref. \cite{BLF}, so that in this approximation the proof
 of the gluon Reggeization really exists.

Let us turn now to the most interesting case - vacuum quantum numbers in
$t$ -channel. From now,  only this case will be considered,  with
additional simplifying restriction of the forward scattering, i.e. $A'=A,
\,\,\,B'=B, \,\, q=0$. Instead of the imaginary part of the forward scattering
amplitude, we'll consider the total cross section $\sigma_{AB}(s)$,
\begin{equation}
\sigma_{AB}(s)=\frac{Im_s{\cal A}_{AB}^{AB}}{s}.
\label{a47}
\end{equation}
From Eqs.(\ref{a29}), (\ref{a39}) we obtain
\[
\sigma_{AB}(s)=\int_{\delta-i\infty }^{\delta+i\infty }\frac{d\omega }{2\pi i}\,
\frac{1}{(2\pi )^{D-2}}
\int d^{D-2}q _{A\perp}d^{D-2}q _{B\perp }
\left(\frac s{s_0}\right) ^\omega \,
\]
\begin{equation}
\times \frac{\Phi _A({\vec q}_A)}{\vec q^{~2}_A}
G_{\omega}({\vec q}_{A}, {\vec q}_{B})
\frac{\Phi _B({-\vec q}_B)}{\vec q^{~2}_B},
\label{a48}
\end{equation}
where
\[
\Phi _A({\vec q}_A)= I^{(0)}_{AA} = \frac{1}{\sqrt{N^2-1}}\sum_{{\tilde A},c}
|\Gamma_{\tilde A A}^{c}|^2;\,\,\, {\vec q}_A= -{\vec p}_{\tilde A},
\]
\begin{equation}
\Phi _B({\vec q}_B)= I^{(0)}_{BB} = \frac{1}{\sqrt{{N^2-1}}}\sum_{{\tilde B},c}
|\Gamma_{\tilde B B}^{c}|^2; \,\,\,{\vec q}_B= -{\vec p}_{\tilde B}
\label{a49}
\end{equation}
and
\begin{equation}
G_\omega (\vec{q}_1,\vec{q}_2)=
\frac{ G^{(0)}_{\omega}({\vec q}_{1}, {\vec q}_{2}; 0)}
{\vec q_{1}^{~2}\vec q_{2}^{~2}}.
\label{a50}[B
\end{equation}
The PPR vertices $\Gamma_{\tilde P' P}^{c}$ are defined in Eqs.(\ref{a7}),
(\ref{a8}) and the Green functions $G^{(R)}_{\omega}({\vec q}_{1}, {\vec q}_{2};
\vec q) $ in Eqs.(\ref{a41}), (\ref{a42}). Let us note that for quarks and
gluons  the impact factors $\Phi _P({\vec q}_P)$ don't depend really on
${\vec q}_P$ in LLA. We indicate this dependence having in mind possible
generalizations to the cases of scattering of other objects and
corrections to LLA. The Green function $G_\omega (\vec{q}_1,\vec{q}_2)$
satisfies the equation
\begin{equation}
\omega \,G_\omega (\vec{q}_1,\vec{q}_2)=\delta
^{D-2}(\vec{q}_1-\vec{q}_2)+\int d^{D-2}%
\widetilde{q}\,\,\,{\cal K}(\vec{q}_1,\vec{\widetilde{q}})
\,G_\omega (\vec{\widetilde{q}},\vec {q}_2)\,
\label{a51}
\end{equation}
where
\begin{equation}
{\cal K}(\vec{q}_1,\vec{q}_2) =
\frac{{\cal K}^{(0)}({\vec q}_1, {\vec q}_2; 0)}{{\vec q}_1^{~2}{\vec q}_2^{~2}}
 =
2\,\omega (-\vec q_1^{~2})\,\delta^{(D-2)}(\vec{q}_1-\vec{q}_2)
+{\cal K}_r(\vec{q}_1, \vec{q}_2)\,.
\label{a52}
\end{equation}
Here  $\omega (-\vec {q}^{~2})$ is the gluon Regge trajectory given by
Eq.(\ref{a6})
and the integral kernel ${\cal K}_r(\vec{q}_1, \vec{q}_2)$, related with the
real particle production, is given by Eq.(\ref{a26}) at $q=0$:
\[
{\cal K}_r(\vec{q}_1, \vec{q}_2)
=\frac{{\cal K}_r^{(0)}(\vec q_1,\vec q_{2};0)}
{{\vec q}_1^{~2}{\vec q}_2^{~2} }
=\frac 1{2(2\pi )^{D-1}(N^2-1){\vec q_1}^{~2}{\vec q_{2}}^{~2}}
\sum_{c_1,c_2,G_1 }|\gamma _{c_1c_2}^{G_1}(q_1,q_{2})|^2
\]
\begin{equation}
=\frac{g^2N}{(2\pi )^{D-1}}\frac{2}{({\vec q}_1 -{\vec q}_2)^2}.
\label{a53}
\end{equation}

Taken separately, the virtual (\ref{a6}) and real (\ref{a53}) contributions
to the kernel (\ref{a52}) lead to infrared singularities. Indeed,
\begin{equation}
\omega^{(1)}(-\vec q^{~2})=-\frac{g^2N\Gamma(1-\epsilon)}
{(4\pi)^{2+\epsilon }} \frac{2}{\epsilon}(\vec q^{~2})^\epsilon,
\label{a54}
\end{equation}
whereas the real contribution (\ref{a53})
gives the term of order $1/\epsilon$ after integration around the point
$\vec{q}_1 -\vec{q}_2 =0$. These singularities cancel each
other in Eq.(\ref{a51}). Let us demonstrate it for the kernel averaged over the
 azimuthal angle between $\vec{q}_1$ and $\vec{q}_2$. In fact, only
the averaged kernel is relevant until we don't consider spin correlations because
in this  case the impact factors entering  Eq.(\ref{a48}) depend only on
squared transverse momenta. It's clear therefore that the high energy
behaviour of cross sections is determined by the averaged kernel.

 Performing expansion in $\epsilon$ and keeping only the terms giving
non-vanishing contributions at $\epsilon \rightarrow 0$ in (\ref{a51}),
we obtain for the averaged kernel
\begin{equation}
\overline{{\cal
K}_r(\vec{q}_1,\vec {q}_2)}=\frac{g^2N}{(2\pi )^{D-1}} \frac{2}{|\vec q_1^{~2}
-\vec q_2^{~2}|} \left( \frac{\left| \vec q_2^{~2}-\vec q_1^{~2}\right|}
{max({\vec
q}_2^{~2},\vec q_1^{~2})}\right)^{2\epsilon}.
\label{a55}
\end{equation}
Instead of the dimensional regularization we can introduce the cut off
$|\vec q_1^{~2} -\vec q_2^{~2}| > \lambda^2$ in the kernel
${\cal K}_r(\vec{q}_1,\vec {q}_2)$ changing
correspondingly the virtual part.  We obtain
\begin{equation}
\overline{{\cal K}(\vec{q}_1,\vec {q}_2)}=\frac{g^2N} {4\pi^{3}}\left(
-2\ln{\frac{\vec q_1^{~2}}{\lambda^2}}\delta(\vec q_1^{~2} - \vec q_2^{~2})
+\frac{\theta\left(|\vec q_1^{~2} -\vec q_2^{~2}|- \lambda^2\right)}
 {|\vec q_1^{~2} -\vec q_2^{~2}|}\right).
\label{a56}
\end{equation}
The representation (\ref{a56}) permits to find easily such form of the kernel
for which the cancellation of the singularities is evident. For this
purpose it is enough to present
\begin{equation}
2\ln{\frac{\vec q_1^{~2}}{\lambda^2}}=\int d\vec q_2^{~2}\frac{\theta\left(|{\vec
q}_1^2 -\vec q_2^{~2}|- \lambda^2\right)} {|\vec q_1^{~2} -\vec q_2^{~2}|}
\phi (\frac{\vec q_2^{~2}}{\vec q_1^{~2}})
\label{a57}
\end{equation}
with $\phi(1)=1$. Evidently, the mentioned form is not unique. The one adopted in \cite{FKL} and used in literature is
\[
\int d^{D-2}q_2\,\,{\cal K}(\vec{q}_1,\vec {q}_2)f(\vec q_2^{~2})=
\frac{N\alpha_s}{\pi}\int d\vec q_2^{~2}\left[\frac{f({\vec
q_2^{~2}})}{|\vec q_2^{~2} - \vec q_1^{~ 2}|}\right.
\]
\begin{equation}
\left. -f(\vec q_1^{~2})
\frac{\vec q_1^{~2}}{\vec q_2^{~2}}(\frac{1} {|\vec q_2^{~2} - \vec
q_1^{~2}|}-\frac{1}{\sqrt{(\vec q_1^{~2})^2 + 4(\vec q_2^{~2})^2}})\right].
\label{a58}
\end{equation}
In the following we'll use another choice:
\begin{equation}
\int d^{D-2}q_2\,\,{\cal K}(\vec{q}_1,\vec {q}_2)f(\vec q_2^{~2})=
\frac{N\alpha_s}{\pi}\int \frac{ d{\vec q}_2^{~2}}{|{\vec q}_2^{~2}
-{\vec q}_1^{~ 2}|}\left[f({\vec q}_2^{~2})-2\frac{\min (\vec q_2^{~2},
\vec q_1^{~ 2})}{(\vec q_2^{~2}+\vec q_1^{~ 2})}f({\vec q}_1^{~2}) \right].
\label{a59}
\end{equation}

Of course, the representations (\ref{a58}) and (\ref{a59}) are equivalent.
Both of them make explicit the scale invariance of the kernel, due to
which its eigenfunctions are powers of $\vec q_2^{~2}$. We'll take them
as $(\vec q_2^{~2})^{\gamma-1}$ and denote the corresponding
eigenvalues as $\frac{N\alpha_s}{\pi}\chi^B(\gamma)$:
\begin{equation}
\int d^{D-2}q_2\,\,{\cal K}(\vec{q}_1,\vec {q}_2)
(\vec q_2^{~2})^{\gamma-1}=
\frac{N\alpha_s}{\pi}\chi^B(\gamma)(\vec q_1^{~2})^{\gamma-1},
\label{a60}
\end{equation}
so that \cite{FKL}
\begin{equation}
\chi^B(\gamma)=2\psi (1)-\psi (\gamma )-\psi (1-\gamma )\,,\,\,\,\psi
 (\gamma )=\Gamma ^{\prime }(\gamma )/\Gamma (\gamma )\,.
\label{a61}
\end{equation}
The set of functions $(\vec q_2^{~2})^{\gamma-1}$ with $\gamma =1/2+i\nu,
\,\,\, -\infty <\nu<\infty $ is complete, so that we have
\[
\sigma_{AB}(s)=\int_{\delta-i\infty }^{\delta+i\infty }
\frac{d\omega }{2\pi i}\,\int_{-\infty }^{+\infty }
\frac{d\nu }{2\pi^2\left(\omega -\frac{N\alpha_s}{\pi}\chi^B(1/2+i\nu)\right)}
\]
\begin{equation}
\times\int\frac{d^{2}q _{A\perp}}{2\pi}\int\frac{d^{2}q _{B\perp }}
{2\pi}\left(\frac{s}{s_0}\right) ^\omega \,\,\,
\Phi _A({\vec q}_A)(\vec q^{~2}_A)^{-i\nu -3/2}
\Phi _B({-\vec q}_B)(\vec q^{~2}_B)^{i\nu -3/2}.
\label{a62}
\end{equation}
The cross section exists only if the
impact factors possess a good infrared behaviour; otherwise it turns into
infinity. In fact, it is
infinite for scattering of colour particles, as it should be, and
finite for colourless ones, because for them
\begin{equation}
\Phi _P({\vec q}_P)\sim \vec q^{~2}_P
\label{a63}
\end{equation}
for small ${\vec q}_P$.
The maximal value of
$\chi^B(\gamma)$ on the integration contour in
Eq.(\ref{a62}) is $\chi(1/2)=4\ln 2$, that corresponds to the maximal
eigenvalue of the kernel $\omega_P^B = 4N(\alpha_s/{\pi})\ln 2$. At
$\omega=\omega_P^B$ the partial wave has the branch point that leads
to the growth (\ref{a1}) of the cross section.

\section{The Next-to-Leading Approximation}

The next-to-leading logarithmic approximation (NLLA) means that all terms of the type
$\alpha[\alpha_s\ln(1/x)]^n$ have to be collected. It was argued in Ref.
\cite{LF} that in this approximation we can use the approach which
coincides in the main features with that used in LLA.
In general, the programme of the calculations is analogous to that in LLA.
The final goal is
the elastic scattering amplitude. It has to be restored from its
$s$- and $u$-channel imaginary parts. The $s$- channel imaginary part
is given by the unitarity relation (\ref{a9}). Evidently,
Eqs.(\ref{a27})-(\ref{a29}) expressing the elastic scattering amplitudes
in terms of their $s$-channel imaginary parts remain unchanged.
 In the multi-Regge kinematics (MRK)
in NLLA, as well as in LLA, only the amplitudes with the gluon quantum
numbers in the channels with momentum transfers $q_i$ do contribute.
It was  mentioned after Eq.(\ref{a29}) that in each order of  perturbation
theory these amplitudes are dominant, because  only for them there is
no cancellation between $s$- and $u$ - channel contributions.
Moreover,for the same reason only in these amplitudes the leading
terms are real, whereas in other amplitudes they are imaginary. Therefore,
the appearance in the r.h.s. of the unitarity relation (\ref{a9}) of amplitudes with
quantum numbers in $t_i$-channels different from the gluon ones leads to loss of at least
two large logarithms and therefore can be ignored in NLLA. Evidently, it
is a peculiar property of NLLA. In the approximations next to NLLA such
amplitudes do contribute.

As before, the key point
in the calculation of the amplitudes contributing in the unitarity
relation is the gluon Reggeization. In  MRK the real parts of the
contributing amplitudes (only these parts are relevant in NLLA because the
LLA amplitudes are real) are presented in the same form
(\ref{a29}) as in LLA. So, in this kinematics the problem is reduced to the
calculation of the two-loop contribution $\omega^{(2)}(t)$ to the
gluon Regge trajectory $\omega(t)$ and the corrections to the real parts
of the PPR-  and RRG- vertices. Let us note here, that the
PPR-vertices, as well as in LLA, enter the expressions for the impact
factors (see Eq.(\ref{a33})), but not  the expression for the kernel
(see Eq.(\ref{a42}),(\ref{a26})), so that the corrections to these vertices
appear at the intermediate steps of the calculations only.
The one-loop corrections to the PPR vertices were
calculated in Refs. \cite{{FL},{FF}}. Though they are necessary for
the calculation of the corrections to trajectory and RRG vertex, they don't
enter the kernel explicitly, therefore I don't present them here.

But in NLLA, contrary to LLA, MRK is not a single  kinematics that
contributes in the unitarity relation (\ref{a9}). Since we have a
possibility to loose one large logarithm (in comparison with LLA),
 the limitation of the strong
ordering (\ref{a13}) in the rapidity space can not be implied more.
Any (but only one) pair of the produced particles can have a fixed (not
increasing with $s$) invariant mass, i.e. components of this pair can
have rapidities of the same order. This kinematics was called \cite{LF}
quasi-multi-Regge kinematics (QMRK). We can  treat this kinematics
including, together with the  one-gluon production, production of more
complicated states in the Reggeon-Reggeon
(RR) collisions, namely  gluon-gluon (GG)
and quark-antiquark $Q\bar Q$ sates,  as well as  production of "excited" states, i.e. sates with larger number of particles,  in the Reggeon-particle (RP) collisions in the
fragmentation region
of one of the initial particles. Therefore, the partial wave (\ref{a28})
can be presented in the same form (\ref{a39}), but with modified impact
factors and RR Green function. In the definition
of the impact factors (\ref{a33}) we have to include radiative
corrections to the PPR vertices and the contribution of the "excited" states
in the fragmentation region. The equation (\ref{a41}) for the  RR Green
function remains unchanged, as well as the representation (\ref{a42}) of
the kernel, but the gluon trajectory has to be taken in the two-loop
approximation:
\begin{equation}  \label{r1}
\omega(t)=\omega^{(1)}(t) + \omega^{(2)}(t),
\end{equation}
and the part, related with the real particle production,
must contain, together with the contribution from the one-gluon production
in the RR collisions, contributions from the two-gluon and
quark-antiquark productions. The one-gluon contribution must be calculated
with the one-loop accuracy, whereas  the two-gluon and
quark-antiquark contributions have to be taken in the Born
approximation. In the following we consider only the case of the forward
scattering and present the part of the kernel related with the real
production  in the form
\begin{equation}  \label{r2}
{\cal K}_{r}(\vec q_1,\vec q_2)= {\cal K}_{RRG}^{one-loop}(\vec
q_1,\vec
q_2) + {\cal K}_{RRGG}^{Born}(\vec q_1,\vec q_2) + {\cal
K}_{RRQ\bar
Q}^{Born}(\vec q_1,\vec q_2).
\end{equation}

Here there is a subtle point. Calculating  the two-gluon production
contribution to the kernel (as well as the  contribution to the impact factor
from the gluon production in the fragmentation region) we meet
divergencies of integrals over invariant masses of the produced
particles
at  upper limits   (let us call such divergencies ultraviolet).
 These divergencies
correspond to the uncertainties   of the lower limits of integrations
over $s_i$ (see Eq. (\ref{a31})) in  MRK, which were not important in LLA,
but can not be ignored in NLLA. Of course, the reason for the divergencies  is
the absence of a natural bound between MRK and QMRK. In order to give
a precise meaning to the  corresponding contributions and to treat them
carefully we introduce an artificial bound (which, of course, disappears
in final results). The discussion of the separation of the MRK and QMRK
contributions presented in subsection 3.3 is based on the paper \cite{FL?}.

 In the next subsections we'll discuss various contributions to the kernel.

\subsection{The Reggeized Gluon Trajectory}

The two-loop corrections to the gluon trajectory were obtained in
Refs. \cite{{FFQ},{FFK}}. They were expressed in terms of discontinuities of
QCD
scattering amplitudes with gluon quantum numbers in $t$-channel
calculated
at large energy $\sqrt{s}$ and fixed momentum transfer $\sqrt{-t}$  in
the
two-loop approximation. The processes of the quark-quark, gluon-gluon
and
quark-gluon elastic scattering were considered. Independently from the
process,
it was obtained for the case of $n_f$ massless quark flavours:

\begin{equation}
\omega^{(2)}(t) =
\frac{g^2t}{(2\pi)^{D-1}}\int\frac{d^{(D-2)}q_1}
{\vec q^{~2}_1(\vec q_1-\vec q)^{~2}}\left[F(\vec q_1,\vec q)-2F(\vec q_1,
\vec q_1)\right] ~,
\label{r3}
\end{equation}
where $t=q^2=-{\vec q}^{~2}$ and
\[
F(\vec q_1,\vec q) =-\frac{g^2N^2{\vec q}^{~2}}{4(2\pi)^{D-1}}
\int\frac{d^{(D-2)}q_2}{\vec q^{~2}_2(\vec q_2-\vec q)^2}
\left[\ln{\left(\frac{\vec q^{~2}}{%
(\vec q_1-\vec q_2)^2}\right)}-2\psi(D-3)\right.
\]
\[
\left.-\psi\left(3-\frac{D}{2}\right)+2\psi\left(\frac{D}{2}
-2\right)+ \psi(1)
+\frac{2}{(D-3)(D-4)}+\frac{D-2}{4(D-1)(D-3)}%
\right]
\]
\begin{equation}
+\frac{2g^2 Nn_f\Gamma\left(2-\frac{D}{2}\right)
\Gamma^2\left(\frac{D}{2}\right)} {(4\pi)^{\frac{D}{2}}\Gamma\left(D\right)}
 \left(\vec q^{~2}\right)^{(\frac{D}{2}-2)}\,.
\label{r4}
\end{equation}
Eqs.(\ref{r3}) and (\ref{r4}) give us a closed expression for the
two-
loop correction to the gluon trajectory. The independence from the
properties
of the scattered particles, which appears as the  result of
remarkable
cancellations among various terms, sets up a stringent test of
the gluon
Reggeization beyond the leading logarithmic approximation.

The two-loop correction to the trajectory contains both ultraviolet
and
infrared divergencies. The former ones can be easily removed by the
charge
renormalization in the total expression for the trajectory. Since
the
trajectory itself must not be renormalized, we have only to use
the
renormalized coupling constant $g_{\mu}$ instead of the bare one $g$
. In
the ${\overline {MS}}$ scheme
\begin{equation}
g=g_{\mu}\mu^{-\mbox{\normalsize
$\epsilon$}}\left[1+\left(\frac{11}{3}-
\frac{2}{3}\frac{n_f}{N}\right)\frac{\bar
g_{\mu}^2}{2\epsilon}\right]~,
\label{r5}
\end{equation}
where
\begin{equation}
\bar g_{\mu}^2 = \frac{g_{\mu}^2N\Gamma(1-\epsilon)}{(4\pi)^{2+
\mbox{\normalsize $\epsilon$}}}~,  \label{r6}
\end{equation}
and calculation of the integrals in (\ref{r3}), (\ref{r4}) gives \cite{FFK}:
\[
\omega(t) = -\bar g^2_{\mu}\left(\frac{\vec q^{~2}}{\mu ^2}\right)^
{\epsilon}
\frac{2}{\epsilon}\left\{1+\frac{\bar
g^2_{\mu}}{\epsilon}
\left[\left(\frac{11}{3}-\frac{2}{3}\frac{n_f}{N}%
\right) \left(1-\frac{\pi^2}{6}\epsilon ^2\right)-\right.\right.
\]
\[
\left.\left. \left(\frac{\vec q^{~2}}{\mu ^2}\right)^
{\epsilon}\left(\frac{11}{6}+\left(\frac{\pi^2}{6}-
\frac{67}{18}\right)
\epsilon+\left(\frac{202}{27}-\frac{11\pi^2}{18}-\zeta(3)\right)
\epsilon^2-\right.\right.\right.
\]
\begin{equation}
\left.\left.\left. \frac{n_f}{3N}\left(1-\frac{5}{3}\epsilon+\left(
\frac{28%
}{9}-\frac{\pi^2}{3}\right)\epsilon^2\right)\right)\right]\right\}.
\label{r7}
\end{equation}
The remarkable fact which occurred is the cancellation of the third
order
poles in $\epsilon$ existing in separate contributions to the
gluon part of (\ref{r3}). This cancellation is
very
important for the absence of infrared divergences in the corrections
to the
BFKL equation. As result of this cancellation, the gluon and
quark
contributions to $\omega^{(2)}(t)$ have similar infrared
behaviour.
Moreover, the coefficient of the leading singularity in $\epsilon$
is
proportional to the coefficient of the one-loop $\beta$ function.
This means
that infrared divergences are strongly correlated with ultraviolet
ones. The
correlation is unique in the sense that it provides the independence
of
singular contributions to $\omega(t)$ from $\vec q^{~2}$. Indeed,
expanding Eq. (%
\ref{r7}) we have
\[
\omega(t) = -\bar g^2_{\mu}\left(\frac{2}{\epsilon}+2\ln\left(
\frac{\vec
q^{~2}}{\mu^2}\right)\right)-\bar g^4_{\mu}\left[\left(\frac{11}{3}
-\frac{2%
}{3}\frac{n_f}{N}\right)\left(\frac{1}{\epsilon^2}-\ln^2\left(
\frac{\vec q^{~2}%
}{\mu^2}\right)\right)\right.
\]

\begin{equation}
\left.
+\left(\frac{67}{9}-\frac{\pi^2}{3}-\frac{10}{9}\frac{n_f}{N}\right)
\left(\frac{1}{\epsilon}+2\ln\left(\frac{\vec
q^{~2}}{\mu^2}\right)\right)-
\frac{404}{27}+2\zeta(3)+\frac{56}{27}\frac{n_f}{N}\right]~.
\label{r8}
\end{equation}
Eq. (\ref{r8}) exhibits explicitly all singularities of the
trajectory in
the two-loop approximation and gives its finite part in the limit
$\epsilon
\rightarrow 0$. Let us stress that the independence of the singular
contributions to the trajectory from $\vec q^{~2}$ is necessary for
the cancellation of the infrared divergencies in the BFKL equation.

\subsection{One-Gluon Production Contribution}

The simplest process for extracting the RRG vertex is
production of one gluon with momentum $k_1$ in MRK:
\begin{equation}  \label{G1}
s\,\gg s_{1,2}, \,\,|t_{1,2}|, \,\,\,\, s_1=(k_1+p_{\tilde A})^2,\,\,\,
s_2=(k_1+p_{\tilde B})^2,\,\,\, s_1s_2\simeq s{\vec k_1}^{~2},\,\,\, t_i=
q_i^2 \simeq -{\vec q_i}^{~2}.
\end{equation}

It is worthwhile to remind, that the MRK amplitudes beyond the LLA have a
complicated analytical structure. For the one-particle production amplitude, assuming the Regge behaviour in the
sub-channels $s_1$ and $s_2$, from general requirements of analiticity,
unitarity and crossing symmetry one has (see Refs. \cite{FL,BAR})
\[
{A}_{AB}^{\tilde A \tilde B +1}=\frac s4\Gamma_{\tilde A A}^{c_1}\frac 1{t_1}
T_{c_1c_2}^d\frac 1{t_2}\Gamma_{\tilde B B}^{c_2}
\]
\[
\times \left\{ \left[ \left( \frac{-s_1}{s_R}\right) ^{\omega _1-\omega
_2}+\left( \frac{s_1}{s_R}\right) ^{\omega _1-\omega _2}\right] \left[
\left( \frac{-s}{s_R}\right) ^{\omega _2}+\left( \frac s{s_R}\right)
^{\omega _2}\right] R\right.
\]

\begin{equation}
\left. +\left[ \left( \frac{-s_2}{s_R}\right) ^{\omega _2-\omega _1}+\left(
\frac{s_2}{s_R}\right) ^{\omega _2-\omega _1}\right] \left[ \left( \frac{-s}{%
s_R}\right) ^{\omega _1}+\left( \frac s{s_R}\right) ^{\omega _1}\right]
L\right\} ,  \label{G2}
\end{equation}
where $ \omega_i= \omega(t_i)$  and  the right and left RRG vertices
$R^{(s_R)}$ and $L^{(s_R)}$ are real
in all physical channels. Fortunately, as it was explained at the begining
of this section, in NLLA  only real parts of the
production amplitudes do contribute , because only these parts interfere with
the LLA amplitudes, which are real. Therefore, for our purposes we can
neglect the imaginary parts and present the amplitudes in the form
({\ref{a17}) with
\[
\gamma _{c_1c_2}^{G_1}(q_1,q_2)=T_{c_1c_2}^d\frac 12\left\{
(R+L)\left( 1-\frac{\omega ^{(1)}(t_1)+\omega ^{(1)}(t_2)}%
2\ln (\frac{{\vec k}_1^2}{s_R})\right) \right.
\]
\begin{equation}
\left. +(R-L)\frac{\omega ^{(1)}(t_1)-\omega ^{(1)}(t_2)}%
2\ln (\frac{{\vec k}_1^2}{s_R})\right\}  \label{G3}
\end{equation}
where $\omega ^{(1)}(t)$ is the one-loop contribution to the gluon Regge
trajectory.

The right and left RRG vertices $R$ and $L$ were calculated in Refs. \cite
{FL}. The calculations were performed in the space-time dimension $D\neq 4$,
but terms vanishing
at $D\rightarrow 4$ were omitted in the final expressions. Unfortunately,
such terms should be kept, because integration over transverse momenta of
the produced gluon leads to divergency at $k_{1\perp }=0$ for the
case $D=4$. Therefore, in the region $k_{1\perp }\rightarrow 0$ we need to know the
production amplitude for arbitrary $\epsilon $. The corresponding
calculations were performed in \cite{FFK1}.

Let us note here, that whereas the dependence of the Regge factors from the
energy scale $s_R$ was beyond the accuracy of LLA, in NLLA it has to be taken
into account, though we can use an arbitrary scale. It means, that the
RRP-vertices, as well as the PPR-vertices become dependent from the energy
scale $s_R$. In the following we'll show  explicitly this dependence,
denoting  the PPG-vertices by $\Gamma _{\tilde P P}^{(s_R)a}$  and the
RRG-vertices by $\gamma_{c_1c_2}^{(s_R)G_1}(q_1,q_2)$. From the physical
requirement of the independence of the production amplitudes from the energy
scale we have:
\begin{equation}
\Gamma _{\tilde P P}^{(s_R)a}=\Gamma_{\tilde P P}^{(s'_R)a}
\left(\frac {s_R}{s'_R}\right)^{\frac{1}{2}\omega(t_{P \tilde P})} \simeq
\Gamma_{\tilde P P}^{(s'_R)a}\left(1+\frac{1}{2}\omega(t_{P\tilde P})
\ln {\left(\frac {s_R}{s'_R}\right)}\right),
\label{G4}
\end{equation}
where  $t_{PP'}$ is the squared momentum transfer from $P$ to $P'$,
and
\[
\gamma_{c_1c_2}^{(s_R)G_1}(q_1,q_2)=\gamma_{c_1c_2}^{(s'_R)G_1}(q_1,q_2)
\left(\frac {s_R}{s'_R}\right)^{\frac{1}{2}(\omega_1+\omega_2)}
\]
\begin{equation}
\simeq\gamma_{c_1c_2}^{(s'_R)G_1}(q_1,q_2)\left(1+\frac{1}{2}(\omega_1+\omega_2)
\ln {\left(\frac {s_R}{s'_R}\right)}\right)
\label{G5}
\end{equation}

In Refs. \cite{FL},\cite{FFK1} the PPR vertices
were extracted from the elastic scattering amplitudes assuming the
representation (\ref{a5}) for
them; therefore, the scales $-t_{P\tilde P}$ were used for these
vertices. In turn, the production vertices $R$ and $L$ were extracted from
the one-gluon production amplitude using  the representation (\ref{G2}),
with the  scale in the Regge factors equal the renormalization scale
$\mu ^2$, but with the PPR vertices taken for the scales
$-t_{P\tilde P}$. So, several scales were mixed in the  one-gluon
production amplitude. Though such mixing of scales is not forbidden,
it is quite inconvenient when we consider production of arbitrary number
of gluons. Therefore we'll use for definition of all vertices the
representation (\ref{a17}) with the single  scale $s_R$.
Then, according to Eqs.(\ref{G4}), (\ref{G5}), the vertices {{$\gamma
_{c_1c_2}^{(s_R)G_1}(q_1,q_2)$}} differ from given by
Eq.(\ref{G3}) with $s_R=\mu ^2$ and vertices $R$ and $L$ taken from
Refs. \cite{FL},\cite{FFK1} by their LLA values (\ref{a19}) multiplied by the
factor
\begin{equation}
\frac 12\left( \omega ^{(1)}(t_1)\ln \left( \frac{-t_1s_R}{\mu ^4}\right)
+\omega ^{(1)}(t_2)\ln \left( \frac{-t_2s_R}{\mu ^4}\right) \right) .
\label{G6}
\end{equation}
Therefore we have
\[
\gamma _{c_1c_2}^{(s_R)G_1}(q_1,q_2)=T_{c_1c_2}^d g_\mu \mu ^{-\epsilon
}e_{\nu} ^{*}(k_1)\Biggl(C^\nu (q_2,q_1)\Biggl[1+\frac{\omega
^{(1)}(t_1)+\omega ^{(1)}(t_2)}2\ln {\left( \frac{s_R}{{\vec k}_1^2}\right)}
\Biggr.\Biggr.
\]
\[
\Biggl.\Biggl.+{\bar g}_\mu ^2\Biggl[(\frac{11}6-\frac{n_f}{3N})
\left( \frac 1\epsilon
+\frac{(\vec q_1^{~2}+\vec q_2^{~2})}{(\vec q_1^{~2}-\vec q_2^{~2})}
\ln {\left(\frac{\vec q_1^{~2}}{\vec q_2^{~2}}\right)}\right)-
\left( \frac{{\vec k}_1^2}{\mu ^2}\right)
^{\epsilon} (\frac 1{\epsilon ^2}-\frac{\pi ^2}2+2\epsilon \zeta (3))
\Biggr.\Biggr.\Biggr.
\]
\[
\Biggl.\Biggl.\Biggl.-\frac 12\ln ^2{\frac{\vec q_1^{~2}}{\vec q_2^{~2}}}
+(1-\frac{n_f}N)\frac{{\vec k}_1^2}3\Biggl(\frac{(\vec q_1^{~2}
+\vec q_2^{~2})}{(\vec q_1^{~2}-\vec q_2^{~2})^2}
-\frac{2\vec q_1^{~2}\vec q_2^{~2}}{(\vec q_1^{~2}
-\vec q_2^{~2})^3}\ln (\frac{\vec q_1^{~2}}{\vec q_2^{~2}})
\Biggr)\Biggr]\Biggr]\Biggl.
\]
\[
\Biggl.+(\frac{p_A}{k_1p_A}-\frac{p_B}{k_1p_B})^\nu \bar g_\mu ^2
\Biggl[(\frac{11}3-\frac{2n_f}{3N})\frac{\vec q_1^{~2}\vec q_2^{~2}}
{(\vec q_1^{~2}-\vec q_2^{~2})}\ln (\frac{\vec q_1^{~2}}{\vec q_2^{~2}})
+(1-\frac{n_f}N)\frac{{\vec k}_{1}^2}3\Biggr.\Biggr.
\]
\begin{equation}
\times \Biggl.\Biggl. \Biggl(\frac{(2{\vec k}_1^2-\vec q_1^{~2}
-\vec q_2^{~2})}{(\vec q_1^{~2}-\vec q_2^{~2})^2}
\left( \frac{\vec q_1^{~2}\vec q_2^{~2}}{(\vec q_1^{~2}-\vec q_2^{~2})}
\ln {(\frac{\vec q_1^{~2}}{\vec q_2^{~2}})}-\frac{(\vec q_1^{~2}
+\vec q_2^{~2})}2\right) +\frac 12\Biggr)\Biggr]\Biggr).
\label{G10}
\end{equation}
As for the vertices $R^{(s_R)}$ and $L^{(s_R)}$ separately, they can be found
taking into account that the combination $R-L$ does not contribute in
the LLA (see (\ref{G3}))  and therefore can be
taken in the first nonvanishing approximation. We obtain
\[
R^{(s_R)}-L^{(s_R)}=R-L,
\]
\[
R^{(s_R)}+L^{(s_R)}=(R+L)\left[ 1+\frac{\omega ^{(1)}(t_1)}2\ln \left( \frac{%
-t_1}{\mu ^2}\right) +\frac{\omega ^{(1)}(t_2)}2\ln \left( \frac{-t_2}{\mu ^2%
}\right) \right]
\]
\begin{equation}
+(R-L)\frac{\omega ^{(1)}(t_1)-\omega ^{(1)}(t_2)}2\ln \left( \frac{s_R}{\mu
^2}\right) ,
 \label{G7}
\end{equation}
where $R$ and $L$ are taken from Refs. \cite{FL},\cite{FFK1} . So, we have
\[
R^{(s_R)}+L^{(s_R)}{=}2g_\mu \mu ^{-\epsilon }e_\nu ^{*}(k_1)\left( C^\nu
(q_2,q_1)\left\{ 1+{\bar g}_\mu ^2\left[ (\frac{11}6-\frac{n_f}{3N})\right.
\right. \right.
\]
\[
\left. \left. \left. \times \left( \frac 1\epsilon +\frac{(\vec q_1^{~2}
+\vec q_2^{~2})}{(\vec q_1^{~2}-\vec q_2^{~2})} \ln (\frac{\vec q_1^{~2}}
{\vec q_2^{~2}})\right)-\left( \frac{{\vec k}_1^2}{\mu ^2}\right) ^\epsilon
\left(\frac 1{\epsilon ^2}-\frac{\pi ^2}2+2\epsilon \zeta (3)-
\epsilon \ln \left( \frac{{\vec k}_1^2}{s_R}\right)
\left( \frac 1{\epsilon ^2}
-\frac{\pi ^2}6\right. \right. \right. \right. \right.
\]
\[
\left. \left. \left. \left. \left. +2\epsilon \zeta (3)\right) \right)
-\frac 12\ln ^2\frac{\vec q_1^{~2}}{\vec q_2^{~2}}+(1-\frac{n_f}N)\frac{{\vec k%
}_1^2}3\left( \frac{(\vec q_1^{~2}+\vec q_2^{~2})}{(\vec q_1^{~2}-{\vec q_2}%
^2)^2}-\frac{2\vec q_1^{~2}\vec q_2^{~2}}{(\vec q_1^{~2}-\vec q_2^{~2})^3}\ln (%
\frac{\vec q_1^{~2}}{\vec q_2^{~2}})\right) \right] \right\} \right.
\]
\[
\left. +(\frac{p_A}{k_1p_A}-\frac{p_B}{k_1p_B})^\nu \bar g_\mu ^2\left[ (%
\frac{11}3-\frac{2n_f}{3N})\frac{\vec q_1^{~2}\vec q_2^{~2}}{(\vec q_1^{~2}-{%
\vec q_2}^2)}\ln (\frac{\vec q_1^{~2}}{\vec q_2^{~2}})+(1-\frac{n_f}N)\frac{{%
\vec k}_{1\perp }^2}3\right. \right.
\]
\begin{equation}
\left. \left. \times \left( \frac{(2{\vec k}_1^2-\vec q_1^{~2}-\vec q_2^{~2})}{%
(\vec q_1^{~2}-\vec q_2^{~2})^2}\left( \frac{\vec q_1^{~2}\vec q_2^{~2}}{({\vec
q_1}^2-\vec q_2^{~2})}\ln (\frac{\vec q_1^{~2}}{\vec q_2^{~2}})-\frac{({\vec q_1%
}^2+\vec q_2^{~2})}2\right) +\frac 12\right) \right] \right) ,  \label{G8}
\end{equation}
\begin{equation}
R^{(s_R)}-L^{(s_R)}=2g_\mu \frac{\mu ^{-\epsilon }e_\nu ^{*}(k)C^\nu
(q_2,q_1)}{\omega ^{(1)}(t_1)-\omega ^{(1)}(t_2)}{\bar g}_\mu ^2\left( \frac{%
{\vec k}_1^2}{\mu ^2}\right) ^\epsilon \left( \frac{-2}\epsilon +\frac{%
\epsilon \pi ^2}3-4\epsilon ^2\zeta (3)\right).  \label{G9}
\end{equation}
One can check that substitution of (\ref{G8}), (\ref{G9}) into (\ref{G3})
gives (\ref{G10}).

It will be shown in the next subsection, that for the appropriate
separation between MRK and QMRK  the contribution
${\cal K}_{RRG}^{one-loop}(\vec q_1,\vec q_2)$ to the kernel is equal to
\begin{equation}
{\cal K}_{RRG}^{one-loop}(\vec q_1,\vec q_2)=\frac 1{2(2\pi )^{D-1}(N^2-1){\vec q_1%
}^2\vec q_2^{~2}}\sum_{c_1, c_2, G_1}|\gamma _{c_1c_2}^{({\vec k}%
_1^2)G_1}(q_1,q_2)|^2. \label{G11}
\end{equation}
From Eqs. (\ref{G10}), (\ref{a20}) we obtain
[B\[
{\cal K}_{RRG}^{one-loop}(\vec q_1,\vec q_2)= \frac{\bar
g^2_{\mu}\mu^{-2{%
\mbox{\normalsize $\epsilon$}}}} {\pi^{1+{\mbox{\normalsize
$\epsilon$}}}
\Gamma(1-\epsilon)}\frac{4}{{\vec k_1^{~2}}}\left(1+\bar
g^2_{\mu}\left[
-2\left(\frac{{\vec k_1^{~2}}}{\mu^2}\right)^{\mbox{\normalsize
$\epsilon$}}
\left(\frac{1}{{\epsilon}^2}-\frac{\pi^2}{2} \right.\right.\right.
\]
\[
\left.\left.\left. +2\epsilon\zeta(3)\right) +\left(\frac{11}{3}
-\frac{2n_f%
}{3N}\right) \frac{1}{\epsilon} + \frac{3\vec k_1^{~2}}{(\vec
q_1^{~2}-\vec
q_2^{~2})} \ln\left(\frac{\vec q_1^{~2}}{\vec q_2^{~2}}\right)
\right.\right.
\]
\[
\left.\left. -\ln^2\left(\frac{\vec q_1^{~2}}{\vec q_2^{~2}}\right) +
\left(1-\frac{n_f}{N}\right)\left(\frac{\vec k^{~2}} {(\vec
q_1^{~2}-\vec
q_2^{~2})}\left(1-\frac{\vec k^{~2} (\vec q_1^{~2}+\vec
q_2^{~2}+4\vec q_1\vec q_2)} {3(\vec q_1^{~2}-\vec q_2^{~2})^2}\right)
\ln\left(\frac{\vec
q_1^{~2}}{\vec q_2^{~2}}\right) \right.\right.\right.
\]
\begin{equation}
\left.\left.\left.- \frac{\vec k^{~2}}{6\vec q_1^{~2}\vec q_2^{~2}}
(\vec
q_1^{~2}+\vec q_2^{~2}+2\vec q_1\vec q_2)+\frac{({\vec
k^{~2}})^2(\vec
q_1^{~2}+\vec q_2^{~2})} {6\vec q_1^{~2}\vec q_2^{~2}(\vec
q_1^{~2}-\vec
q_2^{~2})^2} (\vec q_1^{~2}+\vec q_2^{~2}+4\vec q_1\vec
q_2)\right)\right]\right).  \label{G12}
\end{equation}

\subsection{Separation of the MRK and QMRK contributions}

 Let us rewrite Eq. (\ref{a28}) for the case of the forward scattering:
\begin{equation}
f(\omega ,s_0)_{AB}=\int^{\infty }_{s_0}\frac{ds}s\left( \frac s{s_0}\right)
^{-\omega }\sigma_{AB}(s),
\label{D1}
\end{equation}
using
relation (\ref{a47}) and introducing  $s_0$ as the argument  of the partial
wave because
in NLLA the partial wave becomes dependent on the energy scale.
For the contribution $f^{(n)}_{MRK}(\omega ,s_0)_{AB}$ into
the partial wave coming from production of $n$ gluons in  MRK,
repeating the steps leading to Eq. (\ref{a31}) we obtain
\[
f^{(n)}_{MRK}(\omega ,s_0)_{AB}=\frac{1}{(2\pi )^{D-2}}
\int\left(\prod_{i=1}^{n+1}d^{D-2}q_{i\perp }\frac{ds_i}{s_i}
\left( \frac{s_i}{s_R}\right) ^{2\omega_i}
\left(\frac{s_i}{\sqrt{{\vec k}_{i-1}^{~2}
{\vec k}_i^{~2}}}\right)^{-\omega} \theta(s_i-s_{\Lambda})\right)
\]
\begin{equation}
\times \left(\frac{s_0}{\sqrt{{\vec q}_1^{~2}{\vec q}_{n+1}^{~2}}}\right)
^{\omega}\frac{\Phi^{B+v}_A({\vec q}_1;s_R)}{\vec q_1^{~2}}
\left(\prod_{i=1}^n{\cal K}_{RRG}(\vec q_i,\vec q_{i+1};s_R)\right)
\frac{\Phi^{B+v}
_B(-{\vec q}_{n+1}; s_R)}{\vec q^{~2}_{n+1}}.
\label{D2}
\end{equation}
In order to give  the precise meaning to MRK we introduced here
the parameter $s_{\Lambda}$. By definition,  MRK means that all squared
invariant masses are larger than this fixed (not growing with $s$),
but as large as it would be desired  parameter. The contribution
$f^{(n)}_{MRK}(\omega ,s_0)_{AB}$ depends on this parameter although
we don't indicate this dependence explicitly. The dependence from this parameter
disappears in the final answer for the partial wave.
The impact factors (c.f. (\ref{a49}))
\begin{equation}
\Phi^{B+v}_P({\vec q}_P;s_R)= \frac{1}{\sqrt{N^2-1}}
\sum_{{\tilde P},c}
|\Gamma_{\tilde P P}^{(s_R)c}|^2;\,\,\, {\vec q}_P= -{\vec p}_{\tilde P}
\label{D3}
\end{equation}
acquired the upper index $B+v$ for the denotation that they represent
the sum of the Born (LLA) contribution and the virtual correction, as well as
the additional argument to show their dependence from the scale $s_R$ used
in the Regge factors in Eq.(\ref{a17}). Remind, that besides the presented
contributions, the total NLA impact factors contain the contributions from
the gluon production in the corresponding fragmentation regions as well.
The part of the kernel
\begin{equation}
{\cal K}_{RRG}(\vec{q}_i, \vec{q}_{i+1};s_R)
=\frac 1{2(2\pi )^{D-1}(N^2-1){\vec q_i%
}^{~2}{\vec q_{i+1}}^{~2}}\,\,\,\sum_{c_i,c_{i+1},G_i}
|\gamma _{c_ic_{i+1}}^{(s_R)G_i}(q_i,q_{i+1})|^2.
\label{D4}
\end{equation}
represents the contribution of the gluon production with the one-loop accuracy
and also depends on $s_R$.

It is convenient to change the argument $s_R$ for $\vec k_i^{~2}$ in
${\cal K}_{RRG}(\vec{q}_i, \vec{q}_{i+1};s_R)$  and for $s_0$ in
$\Phi^{B+v}_P({\vec q}_P;s_R)$  using Eqs.(\ref{G4}), (\ref{G5}).
Then after integration over $s_i$  we obtain:
\[
f^{(n)}_{MRK}(\omega ,s_0)_{AB}=\frac{1}{(2\pi )^{D-2}}
\int\left(\prod_{i=1}^{n+1}\frac{d^{D-2}q_{i\perp }}{(\omega -2\omega_i)}
\left(\frac{s_{\Lambda}^2}{{\vec k}_{i-1}^{~2}
{\vec k}_i^{~2}}\right)^{-\frac{(\omega-2\omega_i)}{2}} \right)
\]
\[
\times \left(\frac{s_0}{{\vec q}_1^{~2}}\right)^{\frac{\omega-2\omega_1}{2}}
\left(\frac{s_0}{{\vec q}_{n+1}^{~2}}\right)^{\frac{\omega-2\omega_{n+1}}{2}}
\frac{\Phi^{B+v}_A({\vec q}_1;s_0)}{\vec q_1^{~2}}
\left(\prod_{i=1}^n{\cal K}_{RRG}(\vec q_i,\vec q_{i+1};\vec k_i^{~2})
\right)\frac{\Phi^{B+v}_B(-{\vec q}_{n+1}; s_0)}{\vec q^{~2}_{n+1}}
\]
\[
\simeq \frac{1}{(2\pi )^{D-2}}
\int\left(\prod_{i=1}^{n+1}\frac{d^{D-2}q_{i\perp }}{(\omega -2\omega_i)}\right)
\frac{\Phi^{B+v}_A({\vec q}_1;s_0)}{\vec q_1^{~2}}
\left(\prod_{i=1}^n{\cal K}_{RRG}(\vec q_i,\vec q_{i+1};\vec k_i^{~2})
\right)\frac{\Phi^{B+v}_B(-{\vec q}_{n+1}; s_0)}{\vec q^{~2}_{n+1}}
\]
\begin{equation}
\times\left(1- \sum_{i=2}^n \frac{(\omega -2\omega_i)}{2}
\ln \left(\frac{s_{\Lambda}^2}{{\vec k}_{i-1}^{~2}{\vec k}_i^{~2}}\right)
-\frac{(\omega -2\omega_1)}{2}\ln \left(\frac{s_{\Lambda}^2}
{{\vec k}_1^{~2}s_0}\right)  -\frac{(\omega -2\omega_{n+1})}{2}
\ln \left(\frac{s_{\Lambda}^2}{{\vec k}_n^{~2}s_0}\right)\right).
\label{D5}
\end{equation}
In the last equality we performed the  expansion in $(\omega -2\omega_i)$ and
have taken into account the first terms of the expansion only,
as it should be done in NLLA. Let us note for defineteness that for $n=1$ in
the last factor in Eq.(\ref{D5}) only the two last terms remain, whereas for
$n=0$ the whole factor is equal unity, because in this case in Eq. (\ref{D2})
instead of integrations over partial $s_i$ we have an integration over the
total
$s$ which has to be performed from $s_0$ to $\infty$.

Now let us turn to  QMRK. Pay attention, that since the contribution of
this kinematics is subleading,  we can ignore the dependence from the energy
scale. In the exact analogy with the PPR and RRG vertices let us introduce the
effective vertices $\gamma_{c_ic{i+1}}^{GG}(q_i,q_{i+1})$ and
$\gamma_{c_ic{i+1}}^{Q\bar Q}(q_i,q_{i+1})$ for two-gluon and
quark-antiquark productions in the Reggeon-Reggeon collisions as well as
the vertices $\Gamma^c_{P^* P}$ for  production of the "excited" state $P^*$ in the fragmentation
region of the particle $P$ in the process of scattering of this particle off
the Reggeon, so that the  amplitudes of production of $n+1$ particles in
the QMRK are given by Eq. (\ref{a17}) with one of the "old" vertices changed
for corresponding "new" vertex. With this definition the contribution
$f^{(n+1)}_{QMRK}(\omega)_{AB}
$ of the $n+1$ particle production in
 QMRK is:
\[
f^{(n+1)}_{QMRK}(\omega)_{AB}=\frac{1}{(2\pi )^{D-2}}
\int\left(\prod_{i=1}^{n+1}\frac{d^{D-2}q_{i\perp }}{(\omega -2\omega_i)}\right)
\]
\[
\times
\left\{\frac{\Phi^{(B)}_A({\vec q}_1)}{\vec q_1^{~2}}
\left(\prod_{i=1}^n{\cal K}_r^{(B)}(\vec q_i,\vec q_{i+1})
\right)\frac{\Phi^{(r)}_B(-{\vec q}_{n+1})}{\vec q^{~2}_{n+1}}
+ \frac{\Phi^{(r)}_A({\vec q}_1)}{\vec q_1^{~2}}
\left(\prod_{i=1}^n{\cal K}_r^{(B)}(\vec q_i,\vec q_{i+1})
\right)\frac{\Phi^{(B)}_B(-{\vec q}_{n+1})}{\vec q^{~2}_{n+1}}
\right.
\]
\begin{equation}
\left.+\frac{\Phi^{(B)}_A({\vec q}_1)}{\vec q_1^{~2}}
\sum_{j=1}^n \left(\prod_{i=1}^{j-1}{\cal K}_r^{(B)}(\vec q_i,\vec q_{i+1})
\right){\cal K}_{QMRK}(\vec q_j,\vec q_{j+1})
\left(\prod_{i=j+1}^{n}{\cal K}_r^{(B)}(\vec q_i,\vec q_{i+1})
\right)
\frac{\Phi^{(B)}_B(-{\vec q}_{n+1})}{\vec q^{~2}_{n+1}}
\right\},
\label{D6}
\end{equation}
where $\Phi^{(B)}_P({\vec q}_P)$ and ${\cal K}_r^{(B)}(\vec q_i,\vec q_{i+1})$
denote the Born (LLA) values for the   impact factors
(\ref{a49}) and the part of the LLA kernel related with the real gluon
production (\ref{a53}), $\Phi^{(r)}_P({\vec q}_P)$ means the corrections
to the impact factors due to the production of the "excited" states  in the fragmentation regions and
${\cal K}_{QMRK}(\vec q_j,\vec q_{j+1})$  appears as the contribution to the
kernel from QMRK. The last two values depend on the boundary $s_{\Lambda}$
between  MRK and QMRK, though we don't indicate this dependence.
Taking into account that the phase space  (\ref{a11}) includes
$d^{D-1}k_i/((2\pi)^{D-1}2\epsilon_i)$ and that $d^Dk_i =
dk_i^2d^{D-1}k_i/(2\epsilon_i)$, we have
\[
{\cal K}_{QMRK}(\vec q_j,\vec q_{j+1})=\frac 1{2(N^2-1)\vec q_j^{~2}\vec q_{j+1}^{~2}}\int  d\rho_{P_1P_2}
\theta(s_{\Lambda}-k_j^2)
\]
\begin{equation}
\times \sum_{c_{j},c_{j+1},{P_1,P_2}} |\gamma _{c_{j}c_{j+1}}^{P_1P_2}(q_{j},q_{j+1})|^2.
\label{D7}
\end{equation}
Here the sum is
taken over colours of Reggeons and all
quantum numbers of produced particles. The produced particles $P_1P_2$ can
be quark-antiquark as well
as two gluons; denoting their momenta by $l_1$ and $l_2$  we have
\begin{equation}
d\rho_{P_1P_2} = dk_j^2 \delta (k_j-l_1-l_2) \prod_{ i=1}^2 \frac{d^{D-1}l_i}
{(2\pi)^{D-1}2\epsilon_i}.
\label{D8}
\end{equation}
Note that in the case of the two-gluon production the gluon identity must
be taken into account by the limitation on the integration region.
We can represent Eq. (\ref{D7}) as the following:
\begin{equation}
{\cal K}_{QMRK}(\vec q_j,\vec q_{j+1})=(N^2-1)\int d s_{RR} \frac {2I_{RR}
\sigma_{RR\rightarrow 2}(s_{RR})\theta(s_{\Lambda}-s_{RR})}{(2\pi
)^D2\vec q_j^{~2}\vec q_{j+1}^{~2}},
\label{D9}
\end{equation}
where $\sigma_{RR\rightarrow 2}(k_j^2)$ is the total cross section of the
two-particle production in  collision of Reggeons with momenta $q_j$ and
$-q_{j+1}, \,\,\,\,\,(q_{j}-q_{j+1})^2=s_{RR}$,  averaged over colours of the
Reggeons, and $I_{RR}$ is the invariant flux:
\begin{equation}
I_{RR}=\sqrt{(s_{RR}-q_j^2-q_{j+1}^2)^2-4q_j^2q_{j+1}^2} =
2\sqrt{(q_jq_{j+1})^2-q_j^2q_{j+1}^2};\,\,\,q_i^2=-\vec q_i^{~2}.
\label{D10}
\end{equation}
Analogously,  the correction  $\Phi^{(r)}_P({\vec q}_R)$
to the impact factor due to the production of the "excited" states in the fragmentation
region of the particle $P$ at collision of this particle with momentum $p_P$
and Reggeon with momentum $-q_R,\,\,\, (p_P-q_R)^2=s_{PR}$, is given by
\[
\Phi^{(r)}_P({\vec q}_R)= \frac{(2\pi)^{D-1}}{\sqrt{N^2-1}}
\sum_{P^*,c } \int d\rho_{G\tilde P}
\theta(s_{\Lambda}- s_{PR})|\Gamma^c_{G\tilde A A}|^2
\]
\begin{equation}
=\sqrt{N^2-1}\int d s_{PR} \frac {2I_{PR}
\sigma_{PR\rightarrow P^*}(s_{PR})\theta(s_{\Lambda}-s_{PR})}{(2\pi
)2s}.
\label{D11}
\end{equation}
Here we have taken into account that, with the normalization used the
matrix element for the $RR\rightarrow P_1P_2$ transition coincides with the
corresponding effective vertex, whereas for $PR\rightarrow P^*$
differs from such vertex by factor $\sqrt{2s}$.

Note, that with the NLLA accuracy the corrections obtained by the expansion
in $(\omega-2\omega_i)$ in Eq. (\ref{D5}) for the case of the
$(n+1)$-particle production can be presented in the same form as the r.h.s.
of Eq. (\ref{D6}) with the substitutions:
\[
\frac{\Phi_P^{(r)}({\vec q}_R)}{\vec q_R^{~2}} \rightarrow
-\int d^{D-2}\tilde q_{\perp}
\frac{\Phi^{(B)}_P({\vec {\tilde q}})}{\vec {\tilde q}^{~2}}
{\cal K}_{r}^{(B)}(\vec q_R,\vec {\tilde q})
\frac{1}{2}\ln \left(\frac{s_{\Lambda}^2}{(\vec q_R-\vec {\tilde q})^2s_0}\right),
\]
\begin{equation}
{\cal K}_{QMRK}(\vec q_j,\vec q_{j+1})\rightarrow
 - \int {d^{D-2}\tilde q_{\perp}}
{\cal K}_{r}^{(B)}(\vec q_j,\vec {\tilde q})
{\cal K}_{r}^{(B)}(\vec {\tilde q},\vec q_{j+1})
\frac{1}{2}\ln\left(\frac{s_{\Lambda}^2}{(\vec q_j-\tilde q)^2
(\vec q_{j+1}-\tilde q)^2}\right).
\label{D13a}
\end{equation}
Therefore from Eqs. (\ref{D5})- (\ref{D7}) we obtain that with the NLLA accuracy the
total partial wave can be presented as
\begin{equation}
f(\omega ,s_0)_{AB}=\sum_{n=0}^{\infty}\frac{1}{(2\pi )^{D-2}}
\int\left(\prod_{i=1}^{n+1}\frac{d^{D-2}q_{i\perp }}{(\omega -2\omega_i)}
\right)\frac{\Phi_A({\vec q}_1;s_0)}{\vec q_1^{~2}}
\left(\prod_{i=1}^n{\cal K}_{r}(\vec q_i,\vec q_{i+1})
\right)\frac{\Phi_B(-{\vec q}_{n+1};s_0)}{\vec q^{~2}_{n+1}},
\label{D12}
\end{equation}
where ${\cal K}_{r}(\vec q_1,\vec q_2)$ has the form (\ref{r2}) with
$ {\cal K}_{RRG}^{one-loop}(\vec q_1,\vec q_2)$ given by Eq.(\ref{G11}),
\begin{equation}
{\cal K}_{RRQ\bar Q}^{Born}(\vec q_j,\vec q_{j+1})
 =(N^2-1)\int d s_{RR} \frac {2I_{RR}
\sigma_{RR\rightarrow Q\bar Q}(s_{RR})}
{(2\pi)^D2\vec q_j^{~2}\vec q_{j+1}^{~2}},
\label{D18}
\end{equation}
\[
{\cal K}_{RRGG}^{Born}(\vec q_j,\vec q_{j+1})= (N^2-1)\int d s_{RR}
 \frac {2I_{RR}
\sigma_{RR\rightarrow GG}(s_{RR})\theta(s_{\Lambda}-s_{RR})}
{(2\pi)^D2\vec q_j^{~2}\vec q_{j+1}^{~2}}
\]
\begin{equation}
- \int {d^{D-2}\tilde q_{\perp}}
{\cal K}_{r}^{(B)}(\vec q_j,\vec {\tilde q})
{\cal K}_{r}^{(B)}(\vec {\tilde q},\vec q_{j+1})
\frac{1}{2}\ln \left(\frac{s_{\Lambda}^2}{(\vec q_j-\vec {\tilde q})^2
(\vec q_{j+1}-\vec {\tilde q})^2}\right)\, ,
\label{D14}
\end{equation}
and the impact factors have the representation
\[
\Phi_P({\vec q}_R;s_0)= \sqrt{N^2-1}\int d s_{PR} \frac {2I_{PR}
\sigma_{PR}(s_{PR})\theta(s_{\Lambda}-s_{PR})}{(2\pi)2s}
\]
\begin{equation}
-\int \frac{d^{D-2}\tilde q_{\perp}}{(2\pi)^{D-1}}
\Phi^{(B)}_P({\vec {\tilde q}})\frac{g^2N \vec q_R^{~2}}{\vec {\tilde q}^{~2}
(\vec q_R-\vec {\tilde q})^2}\ln \left(\frac{s_{\Lambda}^2}{(\vec q_R-
\vec {\tilde q})^2s_0}
\right).
\label{D13}
\end{equation}
Here we used the form (\ref{a53}) of ${\cal K}_{r}^{(B)}(\vec q_1, \vec q_2)$.

In the last two equations the MRK boundary $s_\Lambda$ have to be taken much
larger than typical squared transverse momenta, so that the dependence
from $s_\Lambda$ disappears due to the factorization properties
of the Reggeon vertices in the regions of strongly ordered
rapidities of produced particles. Namely, the two-gluon production
vertex is given by
the product of the RRG vertices:
\begin{equation}
\gamma _{c_{j}c_{j+1}}^{G_1G_2}(q_{j},q_{j+1}) \simeq
\gamma _{c_{j}\tilde c}^{G_1}(q_{j},q_{j}-l_1)\frac{1}{(q_{j}-l_1)^2}
\gamma _{\tilde c c_{j+1}}^{G_2}(q_{j}-l_1,q_{j+1})
\label{D15}
\end{equation}
at $(p_Bl_2)\ll(p_Bl_1), \,\,\,(p_Al_1)\ll(p_Al_2)$, where $l_1$ and $l_2$
are momenta of the produced gluons,  and the vertices for
the gluon production in the fragmentation regions are given by the product of
the corresponding vertices without gluon and the RRG vertices:
\begin{equation}
\Gamma^c_{G\tilde A A}\simeq
\Gamma^{\tilde c}_{\tilde A A}\frac{1}{(q_{R}+l)^2}
\gamma _{\tilde c c}^{G}(q_{R}+l,q_R)
\label{D16}
\end{equation}
at $(p_Bl)\ll(p_Bp_{\tilde A}), \,\,\, (p_A p_{\tilde A})\ll(p_Al)$,  $l$
is the gluon momentum, $q_R =p_A-p_{\tilde A}-l$\,;
\begin{equation}
\Gamma^c_{G\tilde B B}\simeq
\Gamma^{\tilde c}_{\tilde B B}\frac{1}{(q_{R}-
l)^2}
\gamma _{c \tilde c}^{G}(q_R, q_R -l)
\label{D17}
\end{equation}
at $(p_Al)\ll(p_Ap_{\tilde B}), \,\,\, (p_B p_{\tilde B})\ll(p_Bl)$\,,
$q_R =p_{\tilde B}- p_B+l$\,.

\subsection{Two-Particle Production Contributions}

Investigation of the two-gluon production contribution was started in
Ref.\cite{LF}; the next step was done in Ref.\cite{FL96}.  The final result
was obtained in Ref.\cite{FK1L} by
calculation of differential cross section of the two-gluon production in
the RR
collisions and integration of this cross section over relative
transverse
and longitudinal momenta of the produced gluons.
In a suitable for integration form we obtain:
$$
{\cal K}_{RRGG}^{Born}(\vec q_1,\vec q_2)=\frac{4\pi g^4\mu^{2\epsilon}N^2}
{(2\pi)^D\vec q_1^{~2}\vec q_2^{~2}}
\int_{\delta_R}^{1-\delta_R}
\frac{dx}{x(1-x)} \int \mu^{-2\epsilon}\frac{d^{D-2}l_1}{(2\pi)^{D-1}}
\biggl\{
\frac{(\vec q_1^{~2})^2(\vec q_2^{~2})^2}{4\tilde t_1\tilde{t_2}
\vec l_1^{~2}\vec l_2^{~2}}
$$
$$
+ \frac{1}{2\tilde t_1\tilde{t_2}}\biggl[ -2\vec q_1^{~2}\vec q_2^{~2}
+ \frac{(1 + \epsilon)}{2}x(1-x)
\left( (\vec k_1^2 - \vec q_2^{~2})^2 + (\vec q_1^{~2})^2 \right)
- 4(1 + \epsilon)x(1-x)
(\vec l_1\vec q_1)^2 \biggr]
$$
$$
+ \frac{x(1-x)\vec q_1^{~2}}{\kappa Z}\biggl[ 2\vec q_2^{~2} +(1+\epsilon)\left( 2(1-x)(\vec l_1\vec q_1)
- x(1-x)\vec q_1^{~2} - \vec l_2^{~2} \right) - \epsilon x(1-x) (\vec k_1^2 - \vec q_2^{~2}) \biggr]
$$
$$
+ \left( \frac{x(1-x)\vec q_1^{~2}}{Z} \right)^2\biggl[ \frac{(1 + \epsilon)}{2} - (3
+ 2\epsilon)x(1-x) \biggr] + \biggl( \frac{x\vec q_1^{~2}\vec q_2^{~2}}{2(1-x)\kappa \vec l_1^{~2}}
- \frac{x^2\vec q_1^{~2}\vec q_2^{~2}}{2\vec l_1^{~2}Z} - \frac{x\vec q_1^{~2}\vec q_2^{~2}}{\vec l_1^{~2}\vec l_2^{~2}} \biggr)
$$
$$
+ \frac{1}{\kappa \tilde t_1}\biggl[ -2(1-x)\vec q_1^{~2}\vec q_2^{~2} + (1 + \epsilon) (1-x) (\vec q_1^{~2})^2
+ \frac{(1 + \epsilon)}{2}x(1-x)\left( 2(1-x)\vec q_1^{~2}(\vec k_1^2 - \vec q_2^{~2})  \right.
$$
$$
\left.-x(\vec q_1^{~2})^2 -x(\vec k_1^2 - \vec q_2^{~2})^2 \right) -2(1 + \epsilon)(2-x)(1-x)\vec q_1^{~2}(\vec l_1\vec q_1)
+ 2(1 + \epsilon)(1-x)\left( (\vec l_1\vec q_1)^2  \right)
$$
$$
+ ((\vec l_1 - x\vec k_1)\vec q_1)^2
+ (1 + \epsilon)\vec q_1^{~2}\vec l_2^{~2} \biggr]
+ \frac{x\vec q_1^{~2}\vec l_2^{~2}\left( (1 + \epsilon)
\vec l_2^{~2}-2\vec q_2^{~2} \right)}{\kappa Z\tilde t_1}
+ \frac{x\vec q_1^{~2}(\vec q_2^{~2})^2}{2\kappa Z\tilde t_1} + \frac
{x\vec q_2^{~2}(\vec q_1^{~2})^2}{2\kappa \tilde t_1\vec l_1^{~2}}
$$
\begin{equation}\label
{T1}
- \frac{(1-x)(2(\vec l_1\vec q_1) - \vec q_1^{~2})\vec q_1^{~2}\vec q_2^{~2}}
{2\tilde t_1\vec l_1^{~2}\vec l_2^{~2}} +(1
+ \epsilon) (1-x)^2 \frac{4(\vec l_1\vec q_1)^2
+ (1-4x)(\vec q_1^{~2})^2}{\tilde t_1^{~2}}
\biggr\},
\end{equation}
where $l_1$ and $l_2$ are momenta of gluons produced in collision of
Reggeons with momenta $q_1$ and $ -q_2$, $l_1+l_2=q_1-q_2=k_1$,
\[
q_1=\beta_1 p_1 +q_{1\perp},\,\,  q_2 =-\alpha_2 p_2
+q_{2\perp},\,\, t_i= {q_i}_{\perp}^2 =-\vec q_i^{~2} ,
\]
\[
l_1=x\beta_1p_1+\frac{\vec l_1^{~2}}{x\beta_1s}p_2 +l_{1\perp}, \,\,
\]
\[
\tilde t_1=(q_1-l_1)^2 =- \frac{1}{x}\left( (\vec l_1 - x\vec q_1)^{~2} +
x(1-x){\vec q_1}^{~2} \right),
\]
$$
\tilde{t_2}=(q_1-l_2)^2 = -\frac{1}{1-x}\left( (\vec l_2 -(1- x)\vec q_1)^{~2} +
x(1-x){\vec q_1}^{~2} \right),
$$
\begin{equation}\label{T2}
\kappa = (l_1 + l_2)^2= \frac{((1-x)\vec l_1 - x\vec l_2)^{~2}}{x(1-x)},\,\,\,
Z = -(1-x){\vec l_1}^{~2} - x{\vec l_2}^{~2}.
\end{equation}
Here we omitted terms which vanish after integtation.
The intermediate parameter $\delta_R$ is introduced in (\ref{T1})  to
exclude divergencies of separate terms.  This parameter
should be  considered as infinitely small. In the total expression
for the integral  the dependence on
$\delta_R$ vanishes.

Note, that taken separately some terms in the r.h.s. of Eq.(\ref{T1})
contain "ultraviolet" divergencies at large $|\vec l_1|$; but these divergencies
are artificial in the sense that they cancel each other, so that the total
integral is convergent at large$|\vec l_1|$. In the infrared region for
$\epsilon \rightarrow 0$ all terms have no more than logarithmic divergency.
All divergencies, "ultraviolet" and infrared,  can be regularized
by the same $\epsilon $. Since in the BFKL equation the kernel
has to be integrated over ${\vec q}_2$, or equivalently, over $\vec k_1 $, and the kernel is
singular at $\vec k_1 =0$, we should be careful in performing the expansion in $\epsilon$ ,
in order to keep all terms which give nonvanishing contribution in the
physical limit $\epsilon\rightarrow 0$ after integration
over $\vec k_1$. The result of the integration in (\ref{T1}) is:
\[
{\cal K}_{RRGG}^{Born}(\vec q_1,\vec q_2)=\frac{4{\bar g}_\mu ^4\mu ^{-2\epsilon }}{%
\pi ^{1+\epsilon }\Gamma (1-\epsilon ){\vec q_1}^{~2}{\vec q_2}^{~2}}%
\Biggl\{
\frac{2{\vec q_1}^{~2}{\vec q_2}^{~2}}{{\vec k_1 }^2}\left( \frac{{\vec
k_1 }^2}{\mu ^2}\right) ^\epsilon \Biggl[ \frac 1{\epsilon ^2}-\frac{11}%
6\frac 1\epsilon -\frac{2\pi ^2}3+\frac{67}{18}+\epsilon \Biggl( \frac{11\pi
^2}{36}-\frac{202}{27}
\]
\[
+9\zeta (3)\Biggr) \Biggr] -\frac{({\vec q_1}^{~2}+{\vec q_2}^{~2})(2{\vec
q_1}^{~2}{\vec q_2}^{~2}-3({\vec q_1}{\vec q_2})^2)}{8{\vec q_1}^{~2}{\vec
q_2}^{~2}}-\Biggl( \frac{11}3\frac{{\vec q_1}^{~2}{\vec q_2}^{~2}}{({\vec q_1%
}^{~2}-{\vec q_2}^{~2})}+\frac{(2{\vec q_1}^{~2}{\vec q_2}^{~2}-3({\vec q_1}{%
\vec q_2})^2)}{16{\vec q_1}^{~2}{\vec q_2}^{~2}}({\vec q_1}^{~2}-{\vec q_2}%
^{~2})\Biggr)
\]
\[
\times \ln \left( \frac{{\vec q_1}^{~2}}{{\vec q_2}^{~2}}\right) -\frac 23%
\frac{{\vec q_1}^{~2}{\vec q_2}^{~2}}{({\vec q_1}^{~2}-{\vec q_2}^{~2})^3}%
\Biggl[ \left( 1-\frac{2({\vec q_1}{\vec q_2})^2}{{\vec q_1}^{~2}{\vec q_2}%
^{~2}}\right) \Biggl( {\vec q_1}^{~4}-{\vec q_2}^{~4}-2{\vec q_1}^{~2}{\vec
q_2}^{~2}\ln \left( \frac{{\vec q_1}^{~2}}{{\vec q_2}^{~2}}\right) \Biggr) +(%
{\vec q_1}{\vec q_2})\Biggl( 2{\vec q_1}^{~2}-
\]
\[
-2{\vec q_2}^{~2}-({\vec q_1}^{~2}+{\vec q_2}^{~2})\ln \left( \frac{{\vec q_1%
}^{~2}}{{\vec q_2}^{~2}}\right) \Biggr) \Biggr] +\frac{2{\vec q_1}^{~2}{\vec
q_2}^{~2}({\vec k_1 }(\vec q_1+\vec q_2))}{{\vec k_1 }^2(\vec q_1+\vec
q_2)^2}\Biggl[ \frac 12\ln \left( \frac{{\vec q_1}^{~2}}{{\vec q_2}^{~2}}%
\right) \ln \left( \frac{{\vec q_1}^{~2}{\vec q_2}^{~2}{\vec k_1 }^4}{({%
\vec q_1}^{~2}+{\vec q_2}^{~2})^4}\right) +
\]
\[
+L\left( -\frac{{\vec q_1}^{~2}}{{\vec q_2}^{~2}}\right) -L\left( -\frac{{%
\vec q_2}^{~2}}{{\vec q_1}^{~2}}\right) \Biggr] -\Biggl[ 4{\vec q_1}^{~2}{%
\vec q_2}^{~2}+\frac{({\vec q_1}^{~2}-{\vec q_2}^{~2})^2}4+(2{\vec q_1}^{~2}{%
\vec q_2}^{~2}-3{\vec q_1}^{~4}-3{\vec q_2}^{~4})\frac{(2{\vec q_1}^{~2}{%
\vec q_2}^{~2}-({\vec q_1}{\vec q_2})^2)}{16{\vec q_1}^{~2}{\vec q_2}^{~2}}%
\Biggr]
\]
\begin{equation}
\times \int_0^\infty \frac{dx}{({\vec q_1}^{~2}+x^2{\vec q_2}^{~2})}\ln
\left| \frac{1+x}{1-x}\right| -{\vec q_1}^{~2}{\vec q_2}^{~2}\left( 1-\frac{(%
{\vec k_1 }(\vec q_1+\vec q_2))^2}{{\vec k_1 }^2(\vec q_1+\vec q_2)^2}%
\right) \left( \int_0^1-\int_1^\infty \right) dz\frac{\ln \left( (z{\vec q_1}%
)^2/{\vec q_2}^{~2}\right) }{(\vec q_2-z\vec q_1)^2}\Biggr\},
\label{T3}
\end{equation}
with
\begin{equation}
\psi (z)=\frac{\Gamma ^{\prime }(z)}{\Gamma (z)};\,\,\,\ \ \ \ \
L(z)=\int_0^z\frac{dt}t\ln (1-t).
\label{T4}
\end{equation}

The r.h.s. of Eq.(\ref{T3}) exhibits several  terms with unphysical
singularities, such as  at   $\vec {q}_1^{~2} = \vec {q}_2^{~2}$ and
$(\vec {q}_1+ \vec {q}_2)^2=0$.
It is not difficult to see, that the only real singularity is at
$\vec k_1^{~2} =0$. All others  are  spurious and singular terms cancel
each other. As for the singularity at $\vec k_1^{~2} =0$,
Eq. (\ref{T3}) contains all terms which give in the
physical case $\epsilon \rightarrow 0$ non-zero contributions after the
subsequent
integration over $ \vec k_1$.
For the terms which are infrared divergent at $D=4$,  the region of
such small $\vec k_1$, that is $\ln(1/\vec k_1^{~2}) \sim 1/{\epsilon}\,\,\,$,
gives essential
contribution in the integral over $ \vec k_1$. Therefore, we can not expand
$ ({\vec k_1}^2/{\mu ^2})^{\epsilon}$ in powers of $\epsilon $ in such terms.
  Note, that the calculated contribution is explicitly symmetrical with respect
to substitution $\vec q_1 \leftrightarrow -\vec q_2$. It is quite a non-trivial
task to notice a reason for this symmetry in the starting expression
(\ref{T1}).
Nevertheless, this symmetry is hidden there. It is a consequence of
the invariance of the expression inside the curly brackets in (\ref{T1}), as
well as the phase space  element $dx/(x(1-x))d\vec {l}_1\,\,$, under  the
"left-right symmetry" transformation \cite{FL96}, i.e. under the substitution:
\begin{equation}
\label{T5}
\vec q_1 \leftrightarrow -\vec q_2,\,\,\,
{\vec l_1} \leftrightarrow {\vec l_2},\,  \, \, \, \,
x \leftrightarrow \frac{x{\vec l_2}^{~2}}
{((1-x){\vec l_1}^{~2} + x{\vec l_2}^{~2})}\, .
\end{equation}

The quark-antiquark pair production contribution was calculated in Refs.
\cite{CCH}.
The result is

\[
{\cal K}_{RRQ\overline{Q}}^{Born}(\vec q_1,\vec q_2) =\frac{{4\bar g}_\mu ^4\mu
^{-2\epsilon }n_f}{\pi ^{1+\epsilon }\Gamma (1-\epsilon )N^3}\Biggl\{ N^2%
\Biggl[ \frac 1{{\vec k_1 }^{~2}}\left( \frac{{\vec k_1 }^{~2}}{\mu ^2}%
\right) ^\epsilon \frac 23\left( \frac 1\epsilon -\frac 53+\epsilon (\frac{28%
}9-\frac{\pi ^2}6)\right)
\]
\[ +\frac 1{(\vec q_1^{~2}-\vec q_2^{~2})}\left[ 1-\frac{{\vec k_1 }^{~2}({%
\vec q_1}^{~2}+\vec q_2^{~2}+4{\vec q_1\vec q_2)}}{3(\vec q_1^{~2}-\vec q_2^{~2})^2%
}\right] \ln \frac{\vec q_1^{~2}}{\vec q_2^{~2}}+\frac{{\vec k_1 }^{~2}}{({%
\vec q_1}^{~2}-\vec q_2^{~2})^2}\left( 2-\frac{{\vec k_1 }^{~2}(\vec q_1^{~2}
+\vec q_2^{~2})}{3\vec q_1^{~2}\vec q_2^{~2}}\right)
\]
\[ +\frac{(2{\vec k_1 }^{~2}-\vec q_1^{~2}-\vec q_2^{~2})}{3\vec q_1^{~2}{\vec
q_2}^{~2}}\Biggr] +\Biggl[ -{1}-\frac{(\vec q_1^{~2}-\vec q_2^{~2})^2}
{4{\vec q_1}%
^{~2}\vec q_2^{~2}}
\]
\[
 +(\frac{({\vec q_1}{\vec q_2})^2}{\vec q_1^{~2}\vec q_2^{~2}}-2)\frac{(2{%
\vec q_1}^{~2}\vec q_2^{~2}-3{\vec q_1}^{~4}-3{\vec q_2}^{~4})}{16\vec q_1^{~2}{\vec q_2%
}^{~2}}\Biggr] \int_0^\infty \frac{dx}{(\vec q_1^{~2}+x^2\vec q_2^{~2})}\ln
\left| \frac{1+x}{1-x}\right|
\]
\begin{equation}
 +\frac{(3({\vec q_1}{\vec q_2})^2-2\vec q_1^{~2}\vec q_2^{~2})}{16{\vec q_1%
}^{~4}{\vec q_2}^{~4}}\left[ (\vec q_1^{~2}-\vec q_2^{~2})\ln \left( \frac{{\vec q_1}%
^{~2}}{\vec q_2^{~2}}\right) +2(\vec q_1^{~2}+\vec q_2^{~2})\right] \Biggr\}.
\label{T6}
\end{equation}

\section{BFKL Pomeron in the next-to-leading approximation}

Now we have all contributions to the NLLA kernel. Remind, that it has the
same form (\ref{a52}) as the LLA one, where the gluon Regge trajectory in
the two-loop approximation is given by (\ref{r8}) and the part
${\cal K}_r(\vec q_1,\vec q_2)$ related with
the real particle production (\ref{r2}) contains the contributions from
the one-gluon (\ref{G12}), two-gluon (\ref{T3}) and quark-antiquark (\ref{T6})
production in the Reggeon-Reggeon collisions.  Representing this part as the sum
\begin{equation}
{\cal K}_r(\vec q_1,\vec q_2)= {\cal K}_r^{(B)}(\vec q_1,\vec q_2)
+{\cal K}_r^{(1)}(\vec q_1,\vec q_2); \,\,\,
{\cal K}_r^{(B)}(\vec q_1,\vec q_2) = \frac{4\overline{g}_\mu
^2\,\mu ^{-2\epsilon }}{\pi ^{1+\epsilon }\,\Gamma (1-\epsilon )}%
\frac 1{(\,{\vec q_1}-{\vec q_2)}^2}
\label{B1}
\end{equation}
of the LLA contribution, expressed in terms of the renormalized coupling
constant, and the one-loop correction ${\cal K}_r^{(1)}(\vec q_1,\vec q_2)$,
we have for this correction \cite{FL98}:
\[
K_r^{(1)}(\vec q_1,\vec q_2)=\frac{4\overline{g}_\mu
^4\,\mu ^{-2\epsilon }}{\pi ^{1+\epsilon }\,\Gamma (1-\epsilon )}%
\left\{ \frac 1{(\,{\vec q_1}-{\vec q_2)}^2}\left[ \left( \frac{11}3-\frac{%
2n_f}{3N}\right) \frac 1\epsilon \left( 1-\left( \frac{(\vec q_1
-\vec q_2)^2}{\mu ^2}\right) ^\epsilon (1-\epsilon ^2%
\frac{\pi ^2}6)\right) \right. \right. \,
\]
\[
\left. \left. +\left( \frac{(\vec q_1-\vec q_2)^2}{%
\mu ^2}\right) ^\epsilon \left( \frac{67}9-\frac{\pi ^2}3-\frac{10}9\frac{%
n_f}{N}+\epsilon \left( -\frac{404}{27}+14\zeta (3)+\frac{56}{27}\frac{%
n_f}{N}\right) \right) \right] \right.
\]
\[
\left. -\left( 1+\frac{n_f}{N^3}\right) \frac{2{\vec q_1}^{~2}{\vec q_2}%
^{~2}-3({\vec q_1}{\vec q_2})^2}{16{\vec q_1}^{~2}{\vec q_2}^{~2}}\left(
\frac 2{{\vec q_2}^{~2}}+\frac 2{{\vec q_1}^{~2}}+(\frac 1{{\vec q_2}%
^{~2}}-\frac 1{{\vec q_1}^{~2}})\ln \frac{{\vec q_1}^{~2}}{{\vec q_2}^{~2}}%
\right) -\frac 1{(\vec q_1-\vec q_2)^2}\left( \ln \frac{{\vec q_1}^{~2}}{{%
\vec q_2}^{~2}}\right) ^2\right.
\]
\[
\left. +\frac{2(\vec  q_1^{~2}-\vec  q_2^{~2})}{(\vec q_1-\vec q_2)^2(\vec q_1+\vec
q_2)^2}\left( \frac 12\ln \left( \frac{{\vec q_1}^{~2}}{{\vec q_2}^{~2}}%
\right) \ln \left( \frac{{\vec q_1}^{~2}{\vec q_2}^{~2}(\vec q_1-\vec q_2)^4%
}{({\vec q_1}^{~2}+{\vec q_2}^{~2})^4}\right) +L\left( -\frac{{\vec q_1}^{~2}%
}{{\vec q_2}^{~2}}\right) -L\left( -\frac{{\vec q_2}^{~2}}{{\vec q_1}^{~2}}%
\right) \right) \right.
\]
\[
\left. -\left( 3+(1+\frac{n_f}{N^3})\left( 1-\frac{({\vec q_1}^{~2}+{\vec
q_2}^{~2})^2}{8{\vec q_1}^{~2}{\vec q_2}^{~2}}-\frac{2{\vec q_1}^{~2}{\vec
q_2}^{~2}-3{\vec q_1}^{~4}-3{\vec q_2}^{~4}}{16{\vec q_1}^{~4}{\vec q_2}^{~4}%
}({\vec q_1}{\vec q_2})^2\right) \right) \int_0^\infty \frac{dx\,\ln \left|
\frac{1+x}{1-x}\right| }{{\vec q_1}^{~2}+x^2{\vec q_2}^{~2}}\right.
\]
\begin{equation}
\left. -\left( 1-\frac{(\vec  q_1^{~2}-\vec  q_2^{~2})^2}{(\vec q_1-\vec q_2)^2(\vec
q_1+\vec q_2)^2}\right) \left( \int_0^1-\int_1^\infty \right) \frac{dz\,\ln
\frac{(z{\vec q_1})^2}{({\vec q_2})^2}}{(\vec q_2-z\vec q_1)^2}\right\} \,,
\label{B2}
\end{equation}
where
\begin{equation}
L(z)=\int_0^z\frac{dt}t\ln (1-t)\,,\,\,\zeta (n)=\sum_{k=1}^\infty k^{-n}.
\label{B3}
\end{equation}
The remarkable fact exhibited by Eq.(\ref{B2}) is the cancellation of the
infrared singularities \cite{F} (remind, that
separate terms in (\ref{r2}) are singular as $1/{\epsilon^2}$)
at fixed $\vec k_1=\vec q_1-\vec q_2$, where we can expand
$ ({\vec k_1}^2/{\mu ^2})^{\epsilon}$ in powers of $\epsilon $. This expansion
is not performed in (\ref{B2}) because for the terms having singularity
at ${\vec k_1}^{~2} =0$  the region of
such small $\vec k_1$, that is $\ln(1/\vec k_1^{~2}) \sim 1/{\epsilon}\,\,\,$,
does contribute in the integral over $ \vec k_1$.  The singular contributions
given by this region are canceled, in turn, in the BFKL equation by the
singular terms in the gluon trajectory \cite{F}.

As well as in LLA, only
the kernel averaged over the angle between the momenta $\vec q_1$ and $%
\vec q_2$
is relevant until we don't consider spin correlations.
For the averaged one-loop correction we have:
\[
\overline{{\cal K}_r^{(1)}(\vec q_1,\vec q_2)}=\frac{4%
\overline{g}_\mu ^4\,\mu ^{-2\epsilon }}{\pi ^{1+\epsilon }\,\Gamma
(1-\epsilon )}\left\{ \frac 1{\left| \vec  q_2^{~2}-\vec  q_1^{~2}\right| }\left[ \left( \frac{%
11}3-\frac{2n_f}{3N}\right) \frac 1\epsilon \left( \left( \frac{\left|
\vec  q_2^{~2}-\vec  q_1^{~2}\right| }{(\max (\vec  q_1^{~2},\vec  q_2^{~2}))}\right) ^{2\epsilon
}(1+\epsilon ^2\frac{\pi ^2}3)\right. \,\right. \right.
\]
\[
\left. \left. \left. -\left( \frac{\left( \vec  q_2^{~2}-\vec  q_1^{~2}\right) ^4}{\mu ^2(\max
(\vec  q_1^{~2},\vec  q_2^{~2}))^3}\right) ^\epsilon \left( 1+\epsilon ^2\frac{5\pi ^2}%
6\right) \right) +\left( \frac{\left| \vec  q_2^{~2}-\vec  q_1^{~2}\right| ^4}{\mu ^2(\max
(\vec  q_1^{~2},\vec  q_2^{~2}))^3}\right) ^\epsilon \left( \frac{67}9-\frac{\pi ^2}3-\frac{%
10}9\frac{n_f}{N}\right. \right. \right.
\]
\[
\left. \left. \left. +\epsilon \left( -\frac{404}{27}+14\zeta (3)+\frac{56%
}{27}\frac{n_f}{N}\right) \right) \right] -\frac 1{32}\left( 1+\frac{n_f}{%
N^3}\right) \left( \frac 2{{\vec  q_2^{~2}}}+\frac 2{{\vec  q_1^{~2}}}+(\frac 1{{\vec  q_2^{~2}}%
}-\frac 1{{\vec  q_1^{~2}}})\ln \frac{{\vec  q_1^{~2}}}{{\vec  q_2^{~2}}}\right) \right.
\]
\[
\left. -\frac 1{|\vec  q_1^{~2}-\vec  q_2^{~2}|}\left( \ln \frac{{\vec  q_1^{~2}}}{{\vec  q_2^{~2}}}\right)
^2-\left( 3+(1+\frac{n_f}{N^3})\left( \frac 34-\frac{(\vec  q_1^{~2}+\vec  q_2^{~2})^2}{%
32\vec  q_1^{~2}\vec  q_2^{~2}}\right) \right) \int_0^\infty \frac{dx}{\vec  q_1^{~2}+x^2\vec  q_2^{~2}}\ln
\left| \frac{1+x}{1-x}\right| \right.
\]
\begin{equation}
\left. +\frac 1{\vec  q_2^{~2}+\vec  q_1^{~2}}\left( \frac{\pi ^2}3-4L(\min (\frac{\vec  q_1^{~2}}{\vec  q_2^{~2}%
},\frac{\vec  q_2^{~2}}{\vec  q_1^{~2}})\right) \right\} \,.  \label{B4}
\end{equation}

Instead of the dimensional regularization we can use the cut off
$ |\vec  q_1^{~2}-\vec  q_2^{~2}| > \lambda^2$ changing
correspondingly the virtual part.
Using Eqs. (\ref{a52}), (\ref{r8}), (\ref{B1}) it is possible to verify
that the averaged
NLLA kernel at $\epsilon \rightarrow 0$ is equivalent to the
expression
\[
\overline{K(\vec q_1,\vec q_2)}=-2\frac{\alpha
_s(\mu ^2)N}{\pi ^2}\left\{ \ln \frac{\vec  q_1^{~2}}{{\lambda }^2}+\frac{\alpha
_s(\mu ^2)N}{4\pi }\left[ \left( \frac{11}3-\frac{2n_f}{3N}\right)
\left( \ln \frac{{\vec q_1}^{~2}}{{\lambda }^2}\ln \frac{{\mu }^2}{{\lambda }^2}+%
\frac{\pi ^2}{12}\right) \right. \right.
\]
\[
\left. \left. +\left( \frac{67}9-\frac{\pi ^2}3-\frac{10}9\frac{n_f}{N}%
\right) \ln \frac{{\vec q_1}^{~2}}{{\lambda }^2}-3\zeta (3)\right] \right\} \delta
(\vec  q_1^{~2}-\vec  q_2^{~2})+\frac{\alpha _s(\mu ^2)N}{\pi ^2}\frac{\theta
(|\vec  q_1^{~2}-\vec  q_2^{~2}|-\lambda ^2)}{|\vec  q_1^{~2}-\vec  q_2^{~2}|}
\]
\[
\times \left\{ 1-\frac{\alpha _s(\mu ^2)N}{4\pi }\left[ \left( \frac{11}3-%
\frac{2n_f}{3N}\right) \ln {\left( \frac{|\vec q_1^{~2}-\vec q_2^{~2}|^2}{\max
(\vec  q_1^{~2},\vec  q_2^{~2})\mu ^2}\right) }-\left( \frac{67}9-\frac{\pi ^2}3-\frac{10}9%
\frac{n_f}{N}\right) \right] \right\}
\]
\[
-\frac{\alpha _s^2(\mu ^2)N^2}{4\pi ^3}\left\{ \frac 1{32}\left( 1+\frac{%
n_f}{N^3}\right) \left( \frac 2{{\vec q_2}^{~2}}+\frac 2{{\vec q_1}^{~2}}
+(\frac 1{{\vec q_2}%
^{~2}}-\frac 1{{\vec q_1}^{~2}})\ln \frac{{\vec q_1}^{~2}}{{\vec q_2}^{~2}}\right) \right.
\]
\[
\left. +\frac 1{|\vec  q_1^{~2}-\vec  q_2^{~2}|}\left(
\ln \frac{{\vec q_1}^{~2}}{{\vec q_2}^{~2}}\right)
^2+\left( 3+(1+\frac{n_f}{N^3})\left( \frac 34-\frac{(\vec  q_1^{~2}+\vec  q_2^{~2})^2}{%
32\vec  q_1^{~2}\vec  q_2^{~2}}\right) \right) \int_0^\infty \frac{dx}{\vec  q_1^{~2}+x^2\vec  q_2^{~2}}\ln
\left| \frac{1+x}{1-x}\right| \right.
\]
\begin{equation}
\left. -\frac 1{\vec  q_2^{~2}+\vec  q_1^{~2}}\left( \frac{\pi ^2}3-4L(\min (\frac{\vec  q_1^{~2}}{\vec  q_2^{~2}%
},\frac{\vec  q_2^{~2}}{\vec  q_1^{~2}}))\right) \right\} \,,  \label{B5}
\end{equation}
in the two-dimensional transverse space, with $\lambda \rightarrow 0$%
. Of course, the dependence from $\lambda $ disappears when the kernel acts
on a function. Moreover, the representation (\ref{B5}) permits to find such
form of the kernel for which this cancellation is evident, just in the same way
as it was done at transition from  (\ref{a56}) to (\ref{a59}). We obtain:
\[
\overline{{\cal K}(\vec q_1,\vec q_2)}=
\frac{\alpha _s(\mu ^2)N}{\pi ^2%
}\int d\vec q^{~2}\frac 1{|\vec q_1^{~2}-\vec q^{~2}|}\left(
\delta (\vec q^{~2}-\vec  q_2^{~2})-2\frac{\min
(\vec q_1^{~2},\vec q^{~2})}{(\vec q_1^{~2}+\vec q^{~2})}
\delta (\vec q_1^{~2}-\vec  q_2^{~2})\right)
\]
\[
\times \left[ 1-\frac{\alpha _s(\mu ^2)N}{4\pi }\left( \left( \frac{11}3-%
\frac{2n_f}{3N}\right) \ln {\left( \frac{|\vec q_1^{~2}-\vec q^{~2}|^2}{\max
(\vec q_1^{~2},\vec q^{~2})\mu ^2}\right) }-\left( \frac{67}9-\frac{\pi ^2}3-\frac{10}9\frac{%
n_f}{N}\right) \right) \right]
\]
\[
-\frac{\alpha _s^2(\mu ^2)N^2}{4\pi ^3}\left[ \frac 1{32}\left( 1+\frac{n_f%
}{N^3}\right) \left( \frac 2{{\vec q_2}^{~2}}+\frac 2{{\vec q_1}^{~2}}+
(\frac 1{{\vec q_2}%
^{~2}}-\frac 1{{\vec q_1}^{~2}})\ln \frac{{\vec q_1}^{~2}}{{\vec q_2}^{~2}}
\right)
+\frac{\left( \ln {(\vec q_1^{~2}/\vec  q_2^{~2})}\right)^2}
{|\vec q_1^{~2}-\vec  q_2^{~2}|} \right.
\]
\[
\left.  +\left( 3+(1+\frac{n_f}{N^3})\left( \frac 34-\frac{(\vec q_1^{~2}+\vec  q_2^{~2})^2}{%
32\vec q_1^{~2}\vec q_2^{~2}}\right) \right) \int_0^\infty \frac{dx}
{\vec q_1^{~2}+x^2\vec  q_2^{~2}}\ln
\left| \frac{1+x}{1-x}\right| \right] -\frac 1{\vec q_2^{~2}+\vec q_1^{~2}}\left( \frac{\pi ^2}3
\right.
\]
\begin{equation}
\left.-4L(\min (\frac{\vec q_1^{~2}}{\vec  q_2^{~2}},%
\frac{\vec  q_2^{~2}}{\vec q_1^{~2}}))\right)
+\frac{\alpha _s^2(\mu ^2)N^2}{4\pi ^3}\left( 6\zeta (3)-\frac{5\pi ^2}{%
12}\left( \frac{11}3-\frac{2n_f}{3N}\right) \right) \delta (\vec q_1^{~2}-\vec  q_2^{~2})\,.
\label{B6}
\end{equation}
The $\mu $ - dependence in the right hand side of this equality leads to the
violation of the scale invariance and is related with running the QCD
coupling constant.

The form (\ref{B6}) is very convenient for finding the action of the kernel
on the eigenfunctions $\vec q_2^{~2(\gamma -1)}$ of the Born kernel:
\begin{equation}
\int d^{D-2}q_2\,\,\,{\cal K}(\vec q_1,\vec q_2)\left(
\frac{\vec  q_2^{~2}}{\vec  q_1^{~2}}\right) ^{\gamma -1}=\frac{\alpha _s(\vec  q_1^{~2})\,N\,}\pi
\left( \chi^B (\gamma )+ \frac{\alpha _s(\vec  q_1^{~2})N}{\pi }
\chi^{(1)} (\gamma )%
\right) \,,\, \label{B7}
\end{equation}
were within our accuracy we expressed the result in terms of the running coupling constant
\begin{equation}
 \alpha _s(\vec q^{~2})=\frac{\alpha _s(\mu ^2)}{1+\frac{\alpha _s(\mu
 ^2)N}{4\pi }%
 \left( \frac{11}3-\frac{2n_f}{3N}\right) \ln {\left( \frac{\vec q^{~2}}{\mu
 ^2}%
 \right) }}\simeq \alpha _s(\mu ^2)\left( 1-\frac{\alpha _s(\mu
 ^2)N}{4\pi }%
 \left( \frac{11}3-\frac{2n_f}{3N}\right) \ln {\left( \frac{\vec q^{~2}}{\mu
 ^2}%
 \right) }\right)\,\, ,
 \label{B8}
 \end{equation}
  $\chi^B(\gamma ) $ is given by  (\ref{a61}) and
 and  the correction $\chi^{(1)}(\gamma)$ is:
\[
\chi^{(1)} (\gamma )=-\frac{1}{4}\left[ \left( \frac{11}3-\frac{2n_f}{3N}\right) \frac
12\left( \chi ^2(\gamma )-\psi ^{\prime }(\gamma )+\psi ^{\prime }(1-\gamma
)\right) \right.
\]
\[
\left. -6\zeta (3)+\frac{\pi ^2\cos(\pi \gamma )}{\sin^2(\pi \gamma
)(1-2\gamma )}\left( 3+\left( 1+\frac{n_f}{N^3}\right) \frac{2+3\gamma
(1-\gamma )}{(3-2\gamma )(1+2\gamma )}\right) \right.
\]
\begin{equation}
\left.-\left( \frac{67}9-\frac{\pi ^2}3-\frac{10}9\frac{n_f}{N}\right)
\chi (\gamma ) -\psi ^{\prime \prime }(\gamma )-\psi ^{\prime \prime }(1-\gamma )-%
\frac{\pi ^3}{\sin (\pi \gamma )}+4\phi (\gamma )\right] \,  \label{B9}
\end{equation}
with
\[
\phi (\gamma )=-\int_0^1\frac{dx}{1+x}\left( x^{\gamma -1}+x^{-\gamma
}\right) \int_x^1\frac{dt}t\ln (1-t)
\]
\begin{equation}
=\sum_{n=0}^\infty (-1)^n\left[ \frac{\psi (n+1+\gamma )-\psi (1)}{(n+\gamma
)^2}+\frac{\psi (n+2-\gamma )-\psi (1)}{(n+1-\gamma )^2}\right] \,.
\label{B10}
\end{equation}

For the relative correction $r(\gamma )$  defined as
 $\chi^{(1)} (\gamma )=-r(\gamma )\chi^{B} (\gamma )\,$
in the symmetrical point $\gamma =1/2$, corresponding to the
eigenfunction of the LLA kernel with the largest eigenvalue,
we have \cite{FL98}
\[
r\left( \frac 12\right) =\left( \frac{11}6-\frac{n_f}{3N}\right) \ln 2-%
\frac{67}{36}+\frac{\pi ^2}{12}+\frac{5}{18}\frac{n_f}{N}+\frac 1{\ln 2}\left[
\int_0^1\arctan (\sqrt{t})\ln (\frac 1{1-t})\frac{dt}t\right.
\]
\begin{equation}
\left. +\frac{11}{8}\zeta (3)+\frac{\pi ^3}{32}\left( \frac{27}{16}+\frac{11}{16}\frac{n_f%
}{N^3}\right) \right] \simeq 6,46 +0.05\frac{n_f}{N}+2.66\frac{n_f}{%
N^3}.  \label{B11}
\end{equation}
It shows that the correction is very large.

In some sense, the large value of the correction is natural
and is a consequence
of the large value of the  LLA Pomeron intercept $\omega_P
^B=4N\ln 2\,\,\alpha _s(q^2)/\pi\,$. If we express the corrected
intercept $\omega_P$
in terms of the Born one,
we obtain
\begin{equation}
\omega_P = \omega_P^B(1-\frac{r\left( \frac 12\right)}{4\ln 2}\omega_P^B)
\simeq \omega_P^B(1-2.4\omega_P^B).
\label{B13}
\end{equation}
The coefficient 2.4 does not look very large. Moreover, it corresponds to
the rapidity interval where correlations become important in
the hadron production processes.

Nevertheless, these numerical estimates show,
that in the kinematical region of HERA
probably it is not enough to take into account only the next-to-leading
correction.
For example, if $\alpha _s(\vec q^{~2})=0.15$, where the Born intercept is
$\omega_P
^B=4N\alpha _s(\vec q^{~2})/\pi \,\ln 2=.39714$, the relative correction for $%
n_f=0 $ is very big:
\begin{equation}
\frac {\omega_P}{\omega_P^B}=1-r\left( \frac 12\right) \,
\frac{\alpha _s(\vec q^{~2})}{%
\pi }N=0.0747.
\label{B12}
\end{equation}
The maximal value of $\omega_P \simeq 0.1$ is obtained
for $\alpha _s(\vec q^{~2})\simeq 0.08$.

But it is necessary to realize that, firstly,
the above estimates are quite straightforward and do not take into account
neither the influence of the running coupling  on the eigenfunctions nor the
nonperturbative effects \cite{L}; secondly, the value of the correction
strongly
depends on its representation.
For example, if one
takes into account the next-to-leading correction by
the corresponding increase of the argument of the running QCD coupling
constant, $\omega_P$ at $\alpha _s(q^2)=0.15$ turns out to be only two
times smaller, than its Born value.

The above results can be applied for the calculation of anomalous dimensions
of the local operators in the vicinity of the point $\omega =J-1=0$.
The deep-inelastic moments ${\cal F}_\omega ^{gB}(q^2)$ defined as
follows
\begin{equation}
{\cal F}_\omega ^{gB}(q^2)=\int_{q^2}^\infty \frac{ds}s\left( \frac
s{q^2}\right) ^{-\omega }\sigma ^{gB}(q^2,s)\,,  \label{B14}
\end{equation}
where
\begin{equation}B
\sigma ^{gB}(q^2,s)=\int \frac{d^2q^{\prime }}{2\pi q^{\prime 2}}\,\Phi _B(%
\vec q^{\prime} )\,\int_{a-i\infty }^{a+i\infty }\frac{d\omega }{%
2\pi i}\,\left( \frac s{q\,q^{\prime }}\right) ^\omega \,G_\omega (%
\vec q,\vec q^{\prime} )\,,  \label{B15}
\end{equation}
obey the integral equation of the type (\ref{a51}) with the inhomogeneous
term equal to $\Phi _B(\vec q)/{(2\pi \vec q^{~2})}$ and the kernel
\begin{equation}
{\widetilde{K}}(\vec q_1,\vec q_2)={\cal K}(\vec q_1,\vec q_2)-
\frac 12\int d^{D-2}{q}\,\,\,{\cal K}^B(%
\vec q_1,\vec q)\,
\ln {\frac{\vec q^{~2}}{\vec q_1^{~2}}}{\cal K}^B(\vec q,\vec q_2)\,.
\label{B16}
\end{equation}
The action of the modified kernel on the Born eigenfunctions $%
\vec q_2^{~2(\gamma -1)}$ can be calculated easily:

\begin{equation}
\int d^{D-2}q_2\,\,\,\widetilde{K}(\vec q_1,\vec q_2%
)\,\,\left( \frac{\vec q_2^{~2}}{\vec q_1^{~2}}\right) ^{\gamma -1}=\frac{\alpha
_s(\vec q_1^{~2})\,N\,}\pi \left( \chi^{B}(\gamma )
+\frac{%
\alpha _s(\vec q_1^{~2})N}{\pi }\widetilde{\chi}^{(1)}(\gamma )\right) \,,\,
\label{B17}
\end{equation}
where
\begin{equation}
\widetilde{\chi}^{(1)}(\gamma )= {\chi}^{(1)}(\gamma )
-\frac{1}{2}\chi (\gamma )\,\chi ^{\prime
}(\gamma )\,.  \label{B18}
\end{equation}
The anomalous dimensions
$\gamma_{\omega}(\alpha_s) =\gamma _0(\alpha _s/\omega )+\alpha _s\gamma _1(\alpha _s/\omega )$
of the twist-2 operators near point $\omega =0$ are determined from the
solution of the equation
\[
\omega =\frac{\alpha _sN}\pi \left( \chi^B (\gamma )
+\frac{\alpha _s N}{\pi }\widetilde{\chi}^{(1)}(\gamma )\right) \simeq
\frac{\alpha _sN}\pi
 \frac 1\gamma  -\frac{\alpha _s^2 N^2}{%
4\,\pi ^2}\left( \frac{11+2n_f/N^3}{3\,\gamma ^2}+\right.
\]
\begin{equation}
\left. \frac{n_f(10+13/N^2)}{9\gamma \,N}+\frac{395}{27}-2\zeta (3)-%
\frac{11}3\frac{\pi ^2}6+\frac{n_f}{N^3}(\frac{71}{27}-\frac{\pi ^2}%
9)\right)   \label{B19}
\end{equation}
for $\gamma \rightarrow 0$. For the low orders of the perturbation theory
we reproduce the
known results and predict the higher loop correction for $\omega \rightarrow
0$:
\[
 \gamma \simeq \frac{\alpha _s\,N}\pi (\frac 1\omega
 -\frac{11}{12}-\frac{n_f}{%
 6N^3})-\left( \frac{\alpha _s\,}\pi \right)
 ^2\frac{n_f\,N}{6\,\omega }%
 (\frac 53+\frac{13}{6N^2})
 \]
 \begin{equation}
 -\frac 1{4\omega ^2}\left( \frac{\alpha _sN}\pi \right) ^3
 \left( \frac{395}{27}-2\zeta (3)-%
 \frac{11}3\frac{\pi ^2}6+\frac{n_f}{N^3}(\frac{71}{27}-\frac{\pi
 ^2}%
 9)\right) \,\,.
\label{B20}
\end{equation}
The results presented in this Section were obtained in Ref.\cite{FL98}.
They differ from the corresponding results of Ref.\cite{CC} because
the results of \cite{CC} were obtained for so called "irreducible part"
of the kernel and therefore are incomplete. After appearance of
Ref.\cite{FL98} the results obtained there were confirmed in
Ref. \cite{CC1}.

Finally, let us estimate the change of the maximal value of the anomalous
dimension $\gamma_{\omega}(\alpha_s)$ which is determined by the position
of the extremum of the function $\omega(\gamma)$ (\ref{B19}). In LLA this
extremum is reached at $\gamma_{max}^B =1/2$ independently from
$\alpha_s$. In  NLLA, assuming the validity
of the perturbative expansion, we obtain
\begin{equation}
\gamma_{max} \simeq \frac{1}{2}-
\frac{N\alpha_s}{\pi}\frac{(\widetilde{\chi}^{(1)}(\frac{1}{2}))^{\prime}}
{(\chi^B(\frac{1}{2}))^{\prime\prime}}=\frac{1}{2}+\frac{N\alpha_s}{4\pi}
(\frac{11}{6}-\frac{n_f}{3N} +8\ln 2).
\label{B21}
\end{equation}
\vskip 1.5cm
{{\bf Acknowledgments}}\\

I thank the Organizing Committee for the financial
support of my participation in the Workshop. The presented results were
obtained in researches supported by the INTAS and RFFI grants.
I am thankful to the Dipartimento di Fisica dell'Universit\`{a} di
Milano, the Dipartimento di Fisica dell'Universit\`{a} della Calabria and
the Istituto Nazionale di Fisica Nucleare - sezione di Milano and gruppo
collegato di Cosenza for their warm hospitality during my stay in Italy.

\end{document}